\documentclass[a4paper]{article}

\usepackage[english]{babel}
\usepackage[latin1]{inputenc}
\usepackage{amssymb,amsfonts,amsmath,amsthm}
\usepackage{a4wide,graphicx,color}
\usepackage{paralist,enumerate}
\usepackage{epstopdf}
\DeclareGraphicsExtensions{.eps}

\usepackage{setspace}
\usepackage{geometry}
\newcommand{\ds}{\displaystyle}
\newcommand{\unit}[1]{~\mathrm{#1}}
\newcommand{\R}{\mathbb{R}}
\newcommand{\X}{\textbf{X}}
\newcommand{\x}{\textbf{x}}
\newcommand{\y}{\textbf{y}}
\newcommand{\dR}{d_{\textup{r}}}
\newcommand{\dA}{d_{\textup{a}}}
\newcommand{\dN}{d_{\textup{n}}}
\newcommand{\FR}{F_{\textup{r}}}
\newcommand{\FA}{F_{\textup{a}}}

\theoremstyle{remark}
\theoremstyle{plain}
	 \newtheorem{proposition}{Proposition}[section]
	 \newtheorem{theorem}[proposition]{Theorem}
	
\theoremstyle{definition}
    \newtheorem{corollary}[proposition]{Corollary}

\begin{document}

\title{Adhesion and volume constraints via nonlocal interactions\\ lead to cell sorting}

\author{Jos\'e Antonio Carrillo \and
        Annachiara Colombi \and
        Marco Scianna
        }


\date{}
	
\maketitle

\begin{abstract}
\medskip
\noindent Abstract.- We demonstrate how concepts of statistical mechanics of interacting particles can have important implications in the choice of interaction potentials to model qualitative properties of cell aggregates in theoretical biology. We illustrate this by showing cell sorting phenomena for cell groups with different adhesiveness parameters: ranging from well-mixed cells aggregates to full segregation of cell type passing through engulfment via adhesiveness tuning.


\medskip
\noindent{\bf Mathematics Subject Classification:}

\end{abstract}

\onehalfspacing

\section*{Introduction}

Adhesive and repulsive cell-cell interactions, and resulting cell patterning, are at the basis of a wide range of biological processes, i.e., from morphogenesis to cancer growth and invasion. For example, defects in the spatial organization of multipotent stem cells in animal embryos lead to severe malformations of adult organs \cite{Ilina3203}. Further, the compact configuration of epithelial monolayers is fundamental in wound healing scenarios. Finally, the dispersion of highly motile malignant individuals triggers the metastatic transition of tumor progression \cite{Friedl2003}.
An accurate analysis of the spatial pattern of cell aggregates is indeed a fundamental issue in theoretical biology. In this respect, apart from extreme situations, most cell systems are characterized by an ordered crystalline structure, where the component cells stabilize at a given minimal distance, larger than nuclei dimensions.

From a mathematical point of view, the patterning of cell aggregates and the intercellular distance can be well described by discrete models. They actually approach the biological problem from a phenomenological point of view, focusing on the cell-level of abstraction and preserving the identity and the behavior of individual elements (for comprehensive reviews the reader is referred to \cite{Alber2003, Anderson2007, Deutsch2007, Drasdo2003}). In more details, these techniques represent biological cells as one or a set of discrete units, while cell morphology is restricted according to some underlying assumptions.
In cell-based methods, the behavior of each individual is then prescribed by a relatively small set of rules, which in general may depend on their phenotype, and on the signals received from the neighbors and/or the environment. However, in this work, we focus on cell dynamics due to intercellular interactions (in particular cell resistance to compression and cell adhesiveness), and neglect both the effect of the environmental conditions as well as any other phenomena (such as, for instance, proliferation and death processes) possibly regulating the evolution of an ensemble of cells.

Some of these mathematical approaches, based on nonlocal interactions between cells forming aggregates, have been already used to model adhesiveness and resistance to compression for cell groups, see \cite{DTGC,GC,PBSG} for instance.
However, a critical discussion of the typical interaction kernels/potentials used in these models and how to find their typical parameters in order to recover qualitative and quantitative real shapes of cell aggregates in applications was lacking. In this respect, we here propose a class of interaction kernels/potentials adapted to reproduce reported behaviors of cell aggregates at the microscopic level such as cell sorting due to cells with different adhesiveness.

One key ingredient to distinguish different cell behaviors is related to a fundamental concept in statistical mechanics of interaction particles called H-stability of the interaction potential, see \cite{ruelle}. This concept was already used in the modelling of swarming or collective behavior of animal populations \cite{OCBC06}, and it turned out to be crucial to determine conditions for large coherent structures of these models such as flocks and mills.

In this context, the main objective of this work is indeed to discuss what are the choices of interaction potentials able to lead crystalline structures and typical distances between cell nuclei (hereafter also called intercellular distances) for microscopic modelling of cell groups. Further, we propose a preliminary strategy to attack the inverse problem of finding well-adapted interaction potentials for different applications in theoretical biology. Finally, we show that this simple model can lead to complicated instabilities if cells with different adhesiveness parameters are involved in the aggregation process. 
More specifically, we observe well-mixed cell groups when heterotypic adhesiveness properties are stronger than homotypic ones; full segregation in the opposite situation (i.e., homotypic adhesiveness properties are stronger than heterotypic ones); and finally partial engulfment of different cell groups for intermediate values of the adhesiveness.

\section{Mathematical model}

Let us consider a biological system composed by $N$ cells of the same phenotype/cell lineage, i.e., characterized by the same biophysical properties (mass and dimension) and behavior, which are distributed over a planar surface. Cells are here represented as discrete entities, i.e., as material particles with concentrated mass $m$ and characterized by the position of their center of mass, $\x_i(t)\in\R^2$ with $i=1,\dots,N$. The configuration of the system at a given instant $t$ is then given by the vector:
\begin{equation*}
    \X(t) = \{\x_1(t), \dots, \x_N(t)\}.
\end{equation*}
\noindent
In order to reproduce cell dynamics arising form intercellular interactions, let us first recall that individual cells in biological environment typically evolve under strong damping, i.e., in an extremely viscous regime (see \cite{SmPl_MMS2012} for a detailed discussion). Hence, we assume that the velocity of individuals and not their acceleration is proportional to the forces applied to the system (i.e., the so-called \textit{overdamped force-velocity response}).
In this respect, the evolution in time of the spatial distribution of the cell aggregate is given by a system of first-order ordinary differential equations. In particular, we further assume that cell dynamics due to direct intercellular interactions results from the superposition of the contributions due to pairwise forces, and consistently avoid cell self-interactions.
Despite cell-cell direct forces may involve several phenomena (such as, for instance, contact-dependent chemical or mechanical interactions), we hereafter take into account only intercellular \emph{repulsive interactions}, which reproduce cell resistance to compression due to size, and \emph{attractive interactions}, which conversely implement cell-cell adhesiveness that rely upon the protrusion and expression of membrane adhesion molecules (e.g., filopodia, cadherins).

Based on these working hypotheses, it is consistent to assume that the velocity contribution arising from pairwise interactions depends on the relative distance between the two cells involved and results aligned toward the line ideally connecting them. The system dynamics is therefore given by
\begin{equation}
    \label{eq:moto-k}
    \dfrac{d\x_i(t)}{dt} = -m\,\sum_{j=1\atop j\neq i}^{N}
    \,K(|\x_i(t) -  \x_j(t)|)
    \,\dfrac{\x_i(t) -  \x_j(t)}{|\x_i(t) -  \x_j(t)|},\quad i=1,\dots,N,
\end{equation}
where $|\cdot|$ identifies the Euclidean norm, and $K:\R_+\rightarrow\R$ (that has units$\unit{\mu m/(\mu g\,s)}$) defines the contribution in the cell velocity due to pairwise interactions. Notice that the cell mass $m$ can be included in the time scale of the system \eqref{eq:moto-k}. However, we prefer to keep it there to possibly compare the effect of variations of $N$ and $m$, both separately (with the consequent increase/decrease of the total mass) and simultaneously (but keeping fixed the total mass of the aggregate).

\begin{figure}[ht]
  \label{fig:1}
  \centering
  \includegraphics[width=1.0\textwidth]{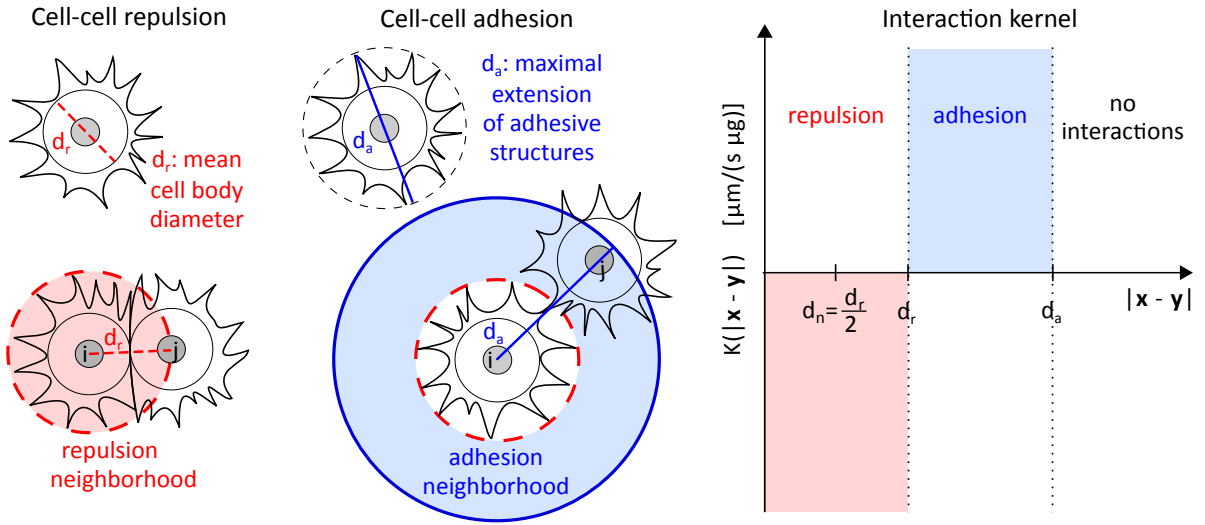}
  \caption{
  Left panel: Repulsive interactions affect the dynamics of two cells when their relative distance is lower than the mean cell body diameter $\dR$. The repulsive neighborhood of each cell $i$, with $i=1,\dots,N$, then consists of a round area with radius $\dR$ centered at the actual position of cell $i$.
  Central panel: Attractive interactions arise between cells whose relative distance is larger than the mean cell body diameter $\dR$ and lower than the maximal extension of cell adhesive molecules $\dA$. Cell adhesive neighborhood then consists of a circular ring with inner radius $\dR$ and external radius $\dA$.
  Right panel: The radial interaction kernel $K$ is assumed negative when repulsive, i.e., $K(|\x-\y|)<0$ when $|\x-\y|<\dR$, and positive when attractive, i.e., $K(|\x - \y|)>0$ if $\dR<|\x-\y|<\dA$, to implement cell resistance to compression and cell-cell adhesiveness, respectively. Further cells whose relative distance is too large do not interact: consistently, we set $K(|\x-\y|)=0$ if $|\x-\y|>\dA$.
  }
\end{figure}

It is obvious from the minimal model \eqref{eq:moto-k} that all the biological mechanisms taken into account are implemented by the shape of the radial interaction kernel $K$. In this respect, we say that $K$ is repulsive when $K(|\x - \y|)<0$ and attractive when $K(|\x - \y|)>0$. On one hand, in order to reproduce cell resistance to compression, it is reasonable to state that cell-cell interactions are repulsive if the relative distance between the interacting cells is lower than the minimal space needed to avoid overlapping, i.e., the mean cell diameter hereafter denoted by $\dR$ (see Fig.~\ref{fig:1}, left panel). On the other hand, as cell-cell adhesiveness rely upon the protrusion of cell filopodia and the expression of cadherins, cell-cell interactions are attractive if the relative distance between two interacting cells is large enough to avoid cell-cell repulsion but small enough to allow the formation of bonds between cell membrane adhesive molecules, but. We thereby assume that $K(|\x - \y|)>0$ if cell relative distance is larger than the mean cell diameter $\dR$ and lower that the maximal extension of cell deformable adhesive structures, that is hereafter denoted by $\dA$ (see Fig.~\ref{fig:1}, central panel). Finally, as we are accounting for cell resistance to compression and cell-cell adhesiveness, thus we assume that if the relative distance between two cells is larger that the maximal extension of membrane adhesive molecules, $\dA$, then $K(|\x - \y|)=0$. This ensures that two cells are not able to mutually interact for distances larger than $\dA$ (see Fig.~\ref{fig:1}, right panel). Similar nonlocal attractive terms to model cell adhesion have already been proposed in the literature of macroscopic and microscopic models of cell interactions for cancer invasion models \cite{DTGC,GC,PBSG} and heterogenous cell populations \cite{BDFS,CHS17,CaSmPl_MMNP2015,CaSmPl_JMB2017,MT15,VS}. In many of these models repulsion is taken into account by (nonlinear) diffusion or drift saturation terms \cite{BDFS,CC,CHS17,Hillen01,Hillen02,SKT} at the macroscopic level. These nonlinear diffusion models can be obtained from microscopic nonlocal repulsive models in the right scaling limit \cite{BV,Ol}, so therefore our choice of repulsive character of the kernel $K$ near the origin.

In principle, there are many options to explicit the form of the interaction kernel $K$ in Eq.~\eqref{eq:moto-k}. As a first step to properly characterize the kernel $K$, let us notice that $K:\R_+\rightarrow\R$ can always be considered as being derived from a scalar interaction potential $u:\R_+\rightarrow\R$ such that $u'(r) = K(r)$, and a vector potential $U:\R^2\rightarrow\R$ such that $U(\x)=u(|\x|)$. Therefore, the right hand side of Eq.~\eqref{eq:moto-k} rewrites in the following form
\begin{equation}
    \label{eq:moto-u}
    \dfrac{d\x_i(t)}{dt} = -m\sum_{j=1}^{N} \nabla U(\x_i(t) - \x_j(t))=
    -m\sum_{j=1}^{N} K\left(|\x_i(t) - \x_j(t)|\right)\dfrac{\x_i(t) - \x_j(t)}{|\x_i(t) - \x_j(t)|},
    \hspace{0.5cm}i=1,\dots,N.
\end{equation}
As already explained, the choice of the interaction kernel/potential is crucial in biological applications as the resulting cell dynamics has to be reliable according to the considered phenomena. In particular, our strategy is to explore the discrete stationary states of system \eqref{eq:moto-k} for selected choices of interaction kernels/potentials, and to analyse the minimal intercellular/interparticle distance
\begin{equation}
    \label{eq:dmin}
    d_{\textup{min}}(t) = \min_{i,j=1,\dots,N\atop i\neq j}|\x_i(t)-\x_j(t)|,
\end{equation}
i.e., the minimal relative distance between cell nuclei. This quantity has a well-defined biological meaning, as required for instance by cell survival or volume exclusion: in this respect, $d_{\textup{min}}(t)$ has to maintain its value above a threshold value (consistently close to the nucleus diameter $\dN$) for any $t$, and will thereby allow us to determine suitable interaction potential shapes. Moreover, in the next sections, we also discuss the scaling properties of stationary states (and the corresponding minimal intercellular distance) depending on $N$ and $m$. In fact, the minimal amount of space needed by each cell to survive is not dependent from both the amount of cells forming the aggregate $N$ and their individual mass $m$. It is conversely given by cell characteristic dimensions as well as by nucleus and cytoplasm stiffness.

\section{Analytical results}
\label{sec:2}

Let us summarize some of the theoretical results concerning the stationary states of \eqref{eq:moto-u}. First of all, notice that this system of equations has the structure of a gradient flow of the total potential energy
$$
E_N(t)=\frac{m}{N^2}\sum_{i,j=1}^N U(\x_i(t)-\x_j(t))\,.
$$
In particular, $E_N$ is a Liapunov functional for \eqref{eq:moto-u} and thus, stable stationary states are among (local or global) minimizers of the interaction energy $E_N$. It is well-known that if the mass of cells scales with the number of particles, i.e., $mN\simeq M$, then the system \eqref{eq:moto-u} converges when $N\to \infty$ to a macroscopic equation, called the aggregation equation, under certain assumptions on the potential and the initial cell configuration, see \cite{BV,meanfield,MVO05,Ol} for instance. This is the so-called mean field limit for interacting particle systems. Let us discuss some of the qualitative properties known for these (local) minimizers.

The first important feature is that the behavior of the solution is regulated by the repulsive part of the interactions, i.e., by the singularity of the potential/kernel at the origin. This fact was studied in \cite{BCLR2,BKSUV} where it was shown that as the potential gets more and more repulsive at the origin the particles distribute in larger and larger regions. In other words, while mild repulsion may allow for clustering of particles, singular repulsion leads to regular distributions of particles in the plane and a well defined minimum interparticle distance.

The second (and more relevant to us) feature is the concept of H-stable potentials introduced in statistical mechanics, see \cite{ruelle}. Assume that the potential is essentially negligible for large distances: i.e., $u$ is such that $\ds\lim_{r\rightarrow +\infty} u(r) = 0 $. Then potentials that are H-stable satisfy that as the number of particles $N\to\infty$, the minimum interparticle distance converges to a fixed value. Therefore, the particles tend to fill the whole plane and the diameter of the particle cloud grows as $N\to\infty$.
While if the potential is not H-stable, sometimes called catastrophic in the statistical mechanics literature, then the diameter of the cloud of particles tend to 0 as $N\to\infty$. Moreover, in the rescaled normalized mass limit $mN\simeq M$ there exists a localized minimizer for the interaction energy. In short, only for H-stable potentials we have a well-defined interparticle distance not depending on $N$ as $N\to\infty$, for any fixed $m$.

This was already pointed out in connection to swarming models for collective behavior of animals in \cite{CDP,OCBC06}, see also \cite{CHM14,CMP13,CHDOB07,HP06}, the reviews \cite{CFTV10,review} and the references therein. The main results about the existence of global minimizers of the interaction energy in the not H-stable case are quite recent \cite{CCP15,CP16,CDM16}. The following easy-to-check criterium to detect H-stability of a potential is already given in \cite{ruelle}, see also \cite{CCP15,CP16} for further discussions and results in the not H-stable cases:
\begin{theorem}\label{thm-loc}
\
    \begin{itemize}
      \item[]If $\ds\int_0^{+\infty}u(r)\,r\,dr>0$, then the potential is \emph{H-stable}.
      \item[]If $\ds\int_0^{+\infty}u(r)\,r\,dr<0$, then there exists a minimizer for the interaction energy in the mean field limit $mN\simeq M$, and the potential is catastrophic or \emph{not H-stable}.
    \end{itemize}
\end{theorem}

\begin{figure}[ht]
  \label{fig:2}
  \centering
  \includegraphics[width=1\textwidth]{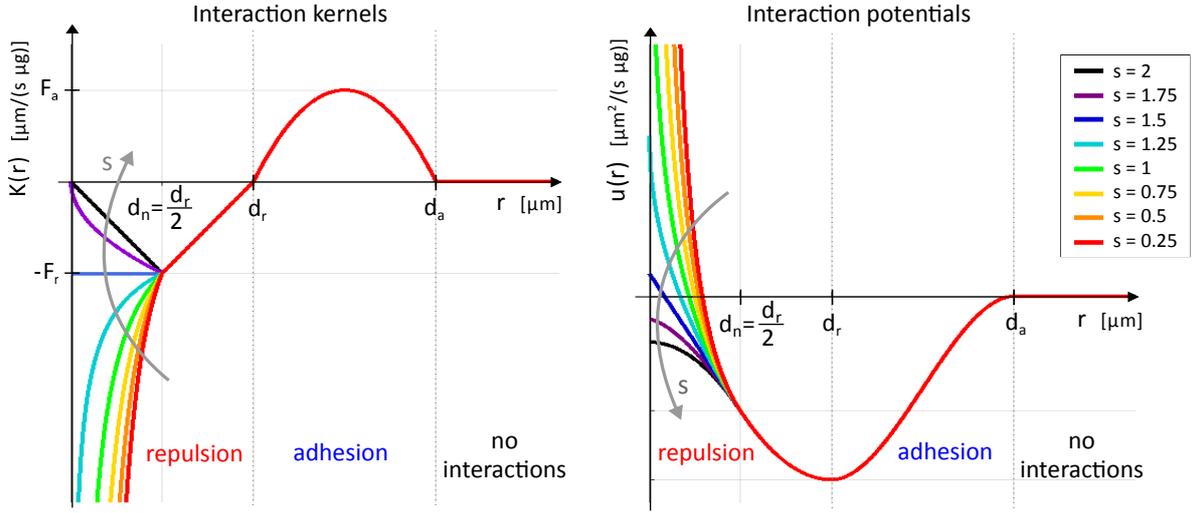}
  \caption{Representation of the set of interaction kernels $K(r)$ defined in Eq.~\eqref{eq:K-s} and the corresponding potentials $u(r)$ in Eq.~\eqref{eq:u-sing}. The interaction kernels present a parabolic behavior in the attractive part and different shapes in the repulsive part characterized by the parameter of $s\in[2,0)$. In particular, we here show the shape of some representative kernels/potentials obtained by setting $s=2,\,1.75,\,1.5,\,1.25,\,1,\,0.75,\,0.5,\,0.25$, we focus on in the following sections.}
\end{figure}

Taking into account the above considerations and theoretical results, we will deal with a family of interaction kernels/potentials and investigate the behavior of a particle system in H-stable and not H-stable conditions to showcase the previous considerations and to choose the right regimes for our purposes. Taking into account that the equilibrium configuration is mainly affected by the singularity of the interaction potential at the origin, we hereafter consider a set of interaction kernels characterized by a parabolic behavior in the attractive part and different shapes in the repulsive part \cite{CaSmPl_JMB2017,CaSmTa_JMB2014}. Specifically, we here introduce a further differentiation of cell repulsive neighborhood in two parts accounting for distinct compressibility of cell nucleus and cytoplasm. In fact, cell nucleus is typically less squeezable than the cytoplasm that is conversely highly deformable and easily compressible. In this respect, let us denote the nucleus diameter by $\dN$ (that is reasonably close to cell radius, i.e., $\dN = \dR/2$), and assume a different explicit form for the interaction kernel depending on whether the intercellular distance is lower than $\dN$ or falls in the range $(\dN, \dR)$. In particular, we hereafter assume $\dN=\dR/2$ and deal with the following set of interaction kernels:
\begin{equation}
    \label{eq:K-s}
    K(r) =
    \left\{
    \begin{array}{ll}
        \ds -\FR\,\left(\dfrac{\dR}{2}\right)^{3-2s}\,r^{2s-3},
            & \hbox{ if } 0 < r \leq \dfrac{\dR}{2}; \\[3mm]
        \ds \dfrac{2\,\FR\,(r-\dR)}{\dR},
            & \hbox{ if } \dfrac{\dR}{2} < r \leq \dR; \\[3mm]
        \ds -\dfrac{4\,\FA\,(r-\dA)\,(r-\dR)}{(\dA-\dR)^2},
            & \hbox{ if } \dR < r \leq \dA; \\[3mm]
        \ds 0,  & \hbox{ if } r > \dA ,
    \end{array}
    \right.
    \hspace{1cm}\textup{with }s\in(0,2],
\end{equation}
where $\FR$ and $\FA$ (that have both units$\unit{\mu m/(\mu g\,s)}$) denote the repulsion and adhesive strengths of cell-cell interactions, respectively; while the parameter $s$ characterizes the kernel behavior at the origin (see Fig.~\ref{fig:2}).

\

\textbf{Interaction kernel with $s\neq1$. - }
From Eq.~\eqref{eq:moto-u}, by setting $s\neq1$, the interaction potential reads as:
\begin{equation*}
    u(r) =
    \left\{
      \begin{array}{ll}
        -\dfrac{\FR}{2}\,\left(\dfrac{\dR}{2}\right)^{3-2s}\dfrac{r^{2s-2}}{(s-1)} +C_1,
            & \hbox{ if } 0 < r \leq \dfrac{\dR}{2};\\[3mm]
        \dfrac{\FR\,r^2}{\dR} -2\,\FR\,r + C_2,
            & \hbox{ if }\dfrac{\dR}{2} < r \leq \dR;\\[3mm]
        -\dfrac{4\,\FA}{(\dA-\dR)^2}
        \,\left(\dfrac{r^3}{3}-\dfrac{(\dA+\dR)\,r^2}{2}+\dA\,\dR\,r\right)+C_3,
            & \hbox{ if } \dR < r \leq \dA;\\[3mm]
        C_4, & \hbox{ if } r > \dA.
      \end{array}
    \right.
\end{equation*}
where the constants of integration $C_1$, $C_2$, $C_3$, $C_4$ $\in\R$ are estimated in order to guarantee the continuity of the potential $u(r)$ and such that $\ds\lim_{r\rightarrow0}u(r)=0$.
Then $C_4 = 0$ and the other constants are set as
\begin{eqnarray*}
    \label{eq:C1}C_1 &=& \dfrac{\FR\,\dR\,s}{4\,(s-1)}-\dfrac{2}{3}\,\FA(\dA-\dR);\\
    \label{eq:C2}C_2 &=& \FR\,\dR - \dfrac{2}{3}\,\FA\,(\dA-\dR);\\
    \label{eq:C3}C_3 &=& \dfrac{2\,\FA\,\dA^2\,(3\,\dR-\dA)}{3\,(\dA-\dR)^2},
\end{eqnarray*}
by continuity at $\dR/2$, $\dR$ and $\dA$.
Thereby the interaction potential writes as
\begin{equation*}
    \label{eq:u-sing}
    u(r) =
    \left\{
      \begin{array}{ll}
        -\dfrac{\FR}{2\,(s-1)}\,\left(\dfrac{\dR}{2}\right)^{3-2s}r^{2s-2} +\dfrac{\FR\,\dR\,s}{4\,(s-1)}-\dfrac{2}{3}\,\FA(\dA-\dR),
            & \hbox{ if } 0 < r \leq \dfrac{\dR}{2};\\[3mm]
        \dfrac{\FR\,r^2}{\dR} -2\,\FR\,r+\FR\,\dR-\dfrac{2}{3}\,\FA\,(\dA-\dR),
            & \hbox{ if }\dfrac{\dR}{2} < r \leq \dR;\\[3mm]
        -\dfrac{2\,\FA\,\left(2\,r^3 - 3\,(\dA+\dR)\,r^2+6\,\dA\,\dR\,r-\dA^2(3\,\dR-\dA)\right)}{3\,(\dA-\dR)^2},
            & \hbox{ if } \dR < r \leq \dA;\\[3mm]
        0, & \hbox{ if } r > \dA.
      \end{array}
    \right.
\end{equation*}
Due to Theorem \ref{thm-loc}, the above interaction potential is H-stable when $\int_0^{+\infty}u(r)\,r\,dr>0$. Taking into account that the values of the interaction radii $\dR$ and $\dA$ are defined according to cell phenotype, the H-stability translates into a constraint on the ratio between the repulsive and adhesive interactions strengths (i.e., $\FR$ and $\FA$) given by
\begin{equation*}
  \begin{array}{l}
    \ds\int_0^{+\infty}u(r)\,r\,dr =
    \dfrac{\FR\,\dR^3\,(11\,s+6)}{192\,s}-\dfrac{\FA}{30}\,(\dA-\dR)\,(3\,\dA^2+4\,\dA\,\dR+3\dR^2)\,>\,0.
    \end{array}
\end{equation*}
Then we have the following Corollary of Theorem \ref{thm-loc}.
\begin{corollary}\label{cor-1}
The interaction potential defined in Eq.~\eqref{eq:K-s}, with $s\neq1$, results H-stable when
\begin{equation}
    \label{eq:F*s}
    \dfrac{\FR}{\FA} > \dfrac{32\,s\,(\dA-\dR)\,(3\,\dA^2+4\,\dA\,\dR+3\dR^2)}{5\,(11\,s+6)\,\dR^3}\,:=\,F^*.
\end{equation}
\end{corollary}.
\

\textbf{Interaction kernel with $s=1$ (hyperbolic case). - }
When the interaction kernel at the origin is characterized by $s=1$, the formulas can be obtained from the previous case with $s\neq1$ by taking the limit as $s\mapsto 1$. This gives
\begin{equation*}
    u(r) =
    \left\{
      \begin{array}{ll}
      -\,\dfrac{\FR\,\dR}{2}\,\log r
      +\dfrac{\FR\,\dR}{2}\,\left(\dfrac{1}{2}+\log \left(\dfrac{\dR}{2}\right)\right)
        -\dfrac{2}{3}\,\FA\,(\dA-\dR),
        & \hbox{ if }0 < r \leq \dfrac{\dR}{2}; \\[3mm]
      \dfrac{\FR\,r^2}{\dR} -2\,\FR\,r+\FR\,\dR-\dfrac{2}{3}\,\FA\,(\dA-\dR),
            & \hbox{ if }\dfrac{\dR}{2} < r \leq \dR;\\[3mm]
        -\dfrac{2\,\FA\,\left(2\,r^3 - 3\,(\dA+\dR)\,r^2+6\,\dA\,\dR\,r-\dA^2(3\,\dR-\dA)\right)}{3\,(\dA-\dR)^2},
            & \hbox{ if } \dR < r \leq \dA;\\[3mm]
        0, & \hbox{ if } r > \dA.
      \end{array}
    \right.
\end{equation*}
In this case, the H-stability Theorem \ref{thm-loc} gives
the constraint presented in the following Corollary.
\begin{corollary}\label{cor-2}
The interaction kernel in Eq.~\eqref{eq:K-s} with $s=1$ is H-stable if
\begin{equation}
    \label{eq:F*1}
    \dfrac{\FR}{\FA} > \dfrac{32\,(\dA-\dR)\,(3\,\dA^2+4\,\dA\,\dR+3\dR^2)}{85\,\dR^3}=F^*.
\end{equation}
\end{corollary}
\noindent
In particular, it is worth to notice that Eq.~\eqref{eq:F*1} also coincides with the relation obtained by setting $s=1$ in Eq.~\eqref{eq:F*s} of Corollary~\ref{cor-1}.

\section{Numerical results}
\label{sec:3}

Accounting for the above theoretical/analytical considerations, this section is devoted to a series of numerical simulations performed to show how the dynamics of pairwise interacting particles are regulated by the H-stability of the interaction kernel and its behavior at the origin. In particular, we here assume that the explicit form of the interaction kernel $K:\R_+\rightarrow\R$ in system \eqref{eq:moto-k} writes as in Eq.~\eqref{eq:K-s} and perform several numerical tests by varying either the value of $s$ (that characterizes the repulsive behavior of $K$ at short distances) or the ratio $\FR/\FA$ between the repulsive and adhesive interaction strength (that conversely defines the H-stability of the interaction kernel).

\bigskip
\noindent
In this perspective, let us first consider a cell aggregate constituted by $N=100$ individuals. In all realizations, cells are initially randomly distributed within a round area of radius equal to $40\unit{ \mu m}$ and centered at the origin, as shown in Fig.~\ref{fig:3} (left panel). In particular, the initial distribution of the aggregate $\X(0)$ is such that for each cell $i=1,\dots,N$ the minimal interparticle distance $d_{\textup{min}}(0)$, defined in Eq.~\eqref{eq:dmin}, is not null (in order to avoid that the center of mass of distinct cells are initially located in the same position) and lower than $\dA$ (so that each cell is initially able to interact at least with another cell). According to \cite{CaSmPl_JMB2017} and the references therein, we hereafter set the cell biophysical properties (mass and size) as follows:
    \begin{center}
    \begin{tabular}{clcc}
      Param. & Description & Value [Unit] & Ref.\\
      \hline
      \noalign{\smallskip}
        $m$ & mean cell mass & $1.8\cdot10^{-3}\unit{ \mu g}$
                & \cite{CaSmPl_MMNP2015}\\
        $d_{\textup{r}}$ & mean cell diameter & $20\unit{ \mu m}$
                & \cite{CaSmPl_MMNP2015}\\
        $d_{\textup{a}}$ & maximal extension of cell filopodia & $60\unit{ \mu m}$
                & \cite{CaSmPl_MMNP2015}\\
      \hline
      \end{tabular}
    \end{center}
\begin{figure}[ht]
    \centerline{\includegraphics[width=1.0\textwidth]{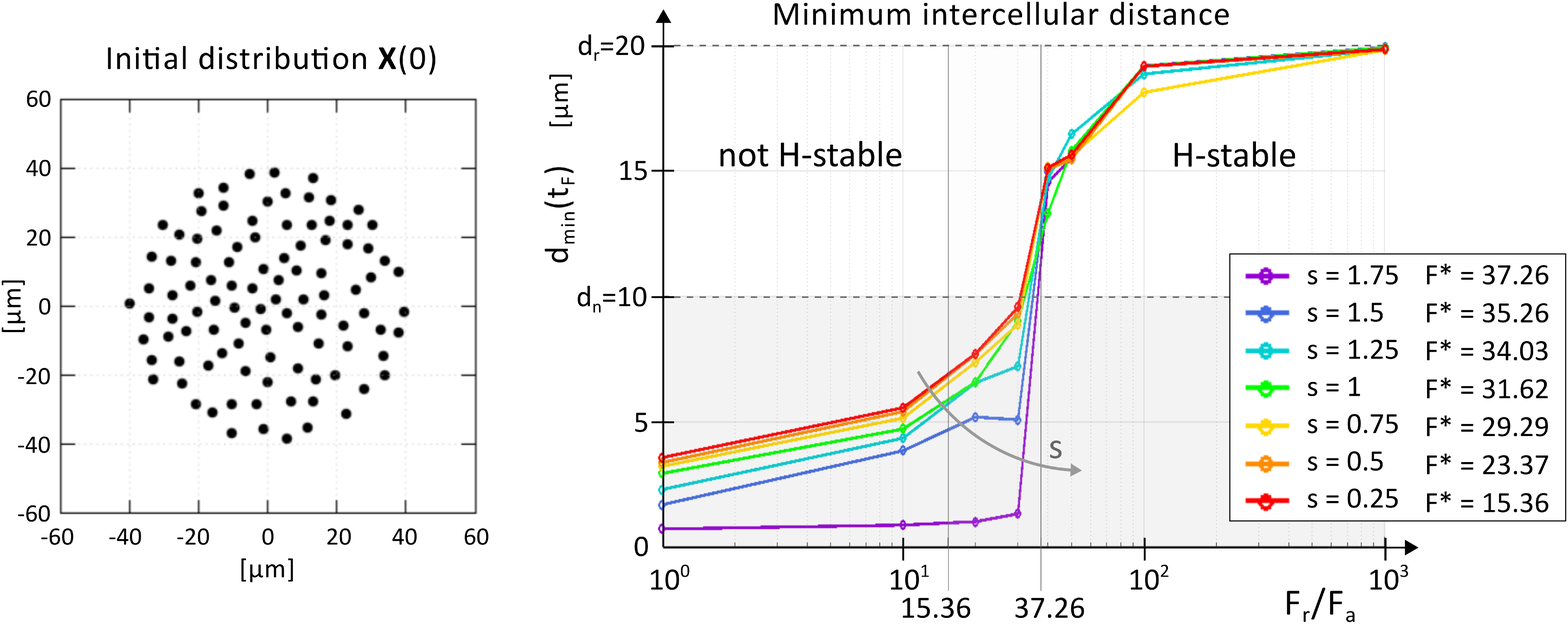}}
    \caption{
    Left panel: Representation of the random spatial distribution of 100 individuals that constitutes the initial condition $\X(0)$ in all realizations of our first series of numerical realizations.
    As a remark, in all simulations, particles are able to move over all the planar space $\R^2$ and no boundary constraints are required.
    Right panel: Variation of the minimum intercellular distance at the equilibrium, i.e., $d_{\textup{min}}(t_{\textup{F}})$ at the equilibrium (see Eq.~\eqref{eq:dmin}), according to different values $\FR/\FA$, observed for distinct choices of $s$, i.e., of the behavior of the pairwise interaction kernel/potential at the origin. 
    According to Theorem~\ref{thm-loc}, the interaction kernel/potential is not H-stable when $\FR/\FA<F^*$ and H-stable when $\FR/\FA>F^*$. For each choice of $s$, the threshold value $F^*$ is reported in the legend.
    The shaded region finally represents unsustainable stationary intercellular distances, i.e., that result in too much packed component cells.
    }
    \label{fig:3}
\end{figure}
From the analytical results provided in the previous section, it thereby follows that the H-stability of the interaction kernel $K$ depends on the ratio of the interaction strengths $\FR$ and $\FA$, according to the value of $s$, i.e., to the shape of the repulsive part of the interaction kernel. Specifically, we here investigate the following cases:
\begin{center}
    \begin{tabular}{c|cccccccc}
        \noalign{\smallskip}
        \hline
        $s$   & 2      & 1.75 & 1.5 & 1.25 & 1 & 0.75 & 0.5 & 0.25 \\
        $F^*$ & 38.40  & 37.26 & 35.26 & 34.02 & 31.32 & 28.29 & 23.37 & 15.36 \\
        \hline
  \end{tabular}
 \end{center}
where $F^*$ is the minimum ratio between the repulsive and adhesive interaction strengths that makes the interaction kernel H-stable (see Eq.~\eqref{eq:F*s} for cases with $s\neq1$ and Eq.~\eqref{eq:F*1} when $s=1$).
In this respect, we first perform a series of numerical simulations exploring the effect of variations of $\FR/\FA$ for each one of the above-cited values of $s$. This allows us to observe how the equilibrium configuration of the system and cell dynamics is regulated by the H-stability of a given interaction kernel/potential.
In more details, in all realizations, we keep the strength of cell-cell adhesiveness $\FA$ constantly equal to $1\unit{ \mu m/(\mu g\,s)}$, so that the value of cell resistance to compression $\FR$ coincides with $\FR/\FA$ and determines the H-stability of $K$. This means that we are fixing the adhesive characteristics of cell-cell interactions, focusing on the repulsive part. However, the discriminating parameter is still the ratio $\FR/\FA$.
In particular, we test all the following values for the repulsive strength: $\FR = 1,\ 10,\ 20,\ 30,\ 40,\ 50,\ 100,\ 1000 \unit{ \mu m/(\mu g s)}$. For all realizations, the evolution of the system has been observed until a stable equilibrium configuration is established and $d_{\textup{min}}(t_F)$ denotes the minimal interparticle distance evaluated at the equilibrium.
The graph in the right panel of Fig.~\ref{fig:3} then shows how the final minimal intercellular distance $d_{\textup{min}}(t_F)$ varies according to the value of the ratio $\FR/\FA$ for each one of considered value of $s$.
According to the above theoretical considerations, it clearly emerges that, independently on the slope of the repulsive part of the kernel (i.e., on $s$), the H-stability of the interaction kernel $K$ regulates the minimum distance between the interacting particles at the equilibrium configuration.
In fact, for any choices of the shape of $K$, we have that:
if $\FR/\FA < F^*$ (i.e., $K$ is not-H stable), the minimal intercellular distance at the equilibrium $d_{\textup{min}}(t_{\textup{F}})$ falls within the range $[0, \dN]$;
while, if $\FR/\FA > F^*$ (i.e., $K$ is H stable), $d_{\textup{min}}(t_{\textup{F}})\in[\dN, \dR]$.
From a biological point of view, these results can be interpreted as it follows:
\begin{itemize}
  \item[-] \emph{not H-stable interaction kernels} (i.e., $\FR/\FA < F^*$) are not able to avoid the superposition of cell nuclei, and thereby the observed equilibrium configurations are biologically unreliable (i.e., $d_{\textup{min}}<\dN$);
  \item[-] conversely, \emph{H-stable interaction kernels} (i.e., $\FR/\FA > F^*$) allow cells to preserve a equilibrium intercellular distance large enough to survive (i.e., $d_{\textup{min}}>\dN$).
\end{itemize}

In this respect, we can further observe that, starting from an initial distribution of cells constituting a single cluster, the minimal intercellular distance at equilibrium will never exceed the repulsion radius $\dR$, i.e., $d_{\textup{min}}(t_{\textup{F}})\leq\dR$, regardless of both the shape and the H-stability of the interaction kernel. In fact, dealing with a first order model where cell dynamics is regulated solely by repulsive-attractive interactions, we have that as $d_{\textup{min}}(t)$ exceeds $\dR$, the repulsive velocity component between any pair of cells vanishes and cells not more drift away.

Referring again to the graph in the right panel of Fig.~\ref{fig:3}, we can moreover notice that in case of not H-stable interaction kernels/potentials the minimal intercellular distance at the equilibrium varies according to the singularity of the repulsive part of $K$. In fact, comparing the results obtained by setting the interaction strengths such that $\FR/\FA < 15.36$ (so that the interaction kernels/potentials are not H-stable for all considered choices of the value of $s$), we have that the value of $d_{\textup{min}}(t_{\textup{F}})$ increases as $s$ decreases, i.e., as $K$ is more singular near to the origin.
On the other hand, we do not observe an analogous correlation between the minimal interparticle distance at the equilibrium and the singularity of the kernel/potential near to the origin in case of H-stable interaction kernels, i.e., when the interaction strength are such that $\FR/\FA > 37.26$ (see again Fig.~\ref{fig:3}, right panel).

\begin{figure}[ht]
    \centerline{\includegraphics[width=1.0\textwidth]{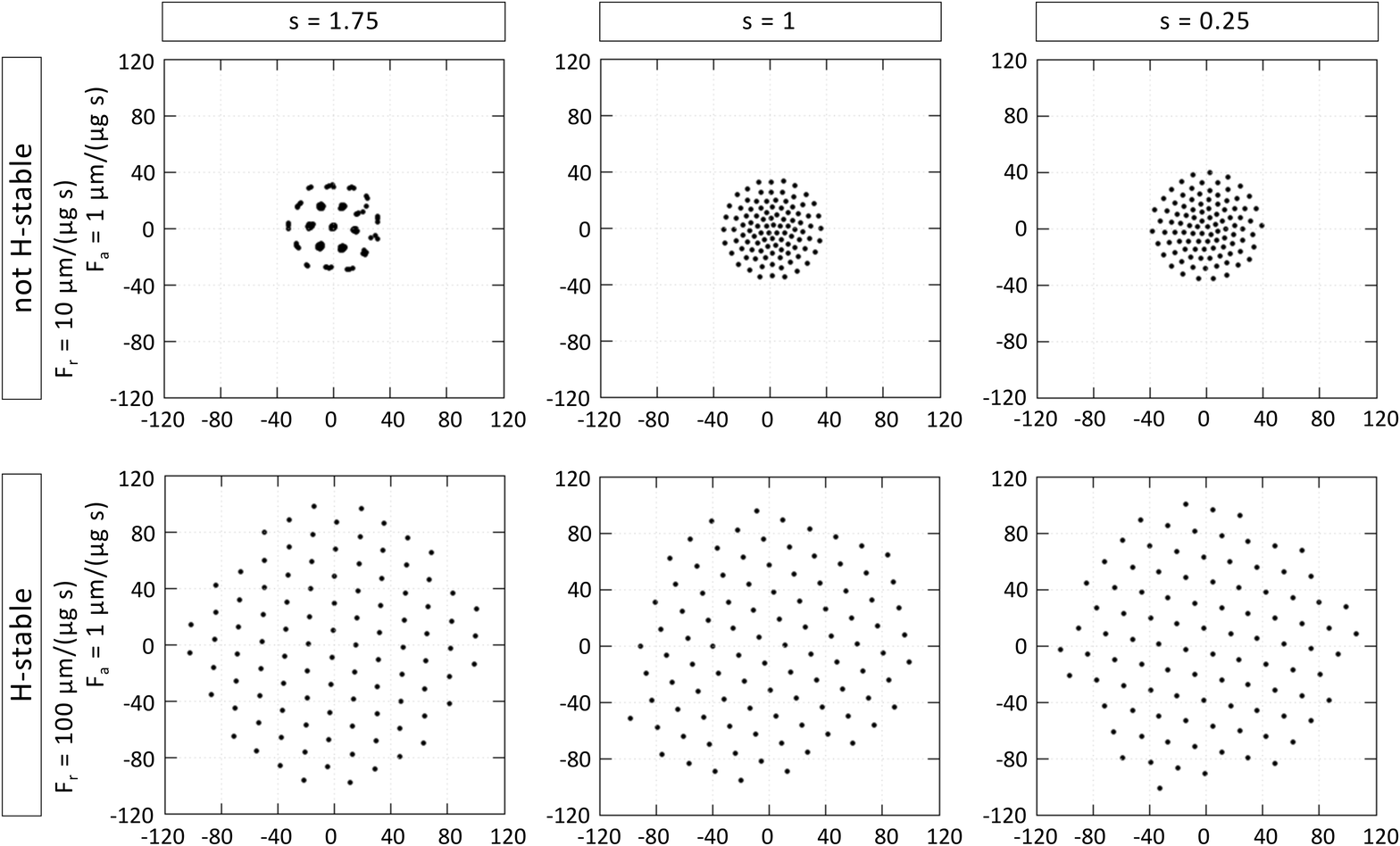}}
    \caption{
    Equilibrium configuration of the cell aggregate (i.e., at $t_F = 8\unit{h}$) observed in the case of some representative choices of both the shape of the interaction kernel at the origin (which is characterized by the value of $s$, i.e., $s = 1.75,\ 1,\ 0.25$), and the ratio between the interaction strengths parameter values (i.e., $\FR/\FA$, which makes the interaction kernels H-stable or not). Specifically, the adhesive strength $\FA$ is here constantly maintained equal to $1\unit{\mu m/(\mu g\,s)}$, so that the value of the ratio $\FR/\FA$ coincides with the value of the repulsive strength $\FR$.
    }
    \label{fig:4}
\end{figure}

In order to further highlight the above commented difference between the not H-stable and H-stable regimes, we report in Fig.~\ref{fig:4} the equilibrium configuration of the aggregate observed in the some representative cases among those described above. Specifically, we hereafter focus only on three possible choices for the behavior of the interaction kernel near the origin (i.e., referring to Eq.~\eqref{eq:K-s}, those corresponding to $s=1.75,\ 1,\ 0.25$), and only to two values of the ratio $\FR/\FA$, i.e., $\FR/\FA = 10$ and $100$, that respectively makes the considered interaction kernels not H-stable and H-stable.
Let us moreover notice that we are comparing the results arising when the interaction kernel at the origin is assumed either more regular than hyperbolic ($s=1.75$), hyperbolic ($s=1$) or more singular than hyperbolic ($s=0.25$).
These final distributions further show that if the interaction kernel is not H-stable (as in the first row in Fig.~\ref{fig:4}), the individuals tend to clusterize and the final minimal intercellular distance depends on the behavior of the interaction kernel at the origin, i.e., on $s$. In particular, we have that by setting $s=1.75$, cells divide into several clusters, with very small intercellular distances within the clusters. Further, the spatial distribution of cells at the equilibrium appears more homogeneous as the value of $s$ is lower, see the cases with $s=1$ and $s=0.25$, i.e., as the interaction kernel/potential is more singular at the origin. However, in all cases, the intercellular distance is always too small to allow cell survival.
On the other hand, dealing with a H-stable interaction kernel (see the second row in Fig.~\ref{fig:4}), cells re-organize in a quite homogeneous distribution characterized by a minimum intercellular distance that allows cell survival. Moreover, such behavior arises for all values of the parameter $s$ here reported as well as those considered above (i.e., $s=0.5,\,0.75,\,1.25\,1.5$).

\bigskip

To investigate deeper how the choice of a not H-stable interaction kernel rather than a H-stable one affects the system dynamics, we focus on the six representative cases reported in Fig.~\ref{fig:4} and analyze the effect of variations of the overall number $N$ and of the individual mass $m$ of cells, both separately (thereby varying the overall mass of the aggregate) and simultaneously by keeping constant the overall mass of the aggregate $M_{\textup{tot}}=mN$.
In more details, for each one of the representative cases considered in Fig.~\ref{fig:4}, we perform a series of numerical simulations to compare the above-commented equilibrium configurations (obtained with $N=100$ cells with mass $m=0.0018 \unit{\mu g}$) with those arising if we deal with:
\begin{itemize}
  \item[(i)] a smaller/larger aggregate of cells whose individual mass $m$ is still equal to $0.0018 \unit{\mu g}$ (specifically, we investigate the cases with $N=50$ and $N=200$ cells);
  \item[(ii)] an aggregate still formed by $N=100$ cells whose individual mass $m$ is lower/higher than $0.0018 \unit{\mu g}$ (in particular, we consider cells with $m=0.0009 \unit{\mu g}$ and $m=0.0036 \unit{\mu g}$, respectively);
  \item[(iii)] an aggregate whose overall mass $M_{\textup{tot}}=m\,N$ is kept equal to $0.18 \unit{\mu g}$, but the number $N$ and, in turn, the individual mass $m$ of cells are different from $100$ and $0.0018 \unit{\mu g}$, respectively, (in this case, we first consider an aggregate composed by $N=50$ cells with $m=0.0036\unit{\mu g}$; then we set $N=200$ and $m=0.0009\unit{\mu g}$). This is related to the mean-field limit as explained in Section~\ref{sec:2}.
\end{itemize}
\begin{figure}[ht]
    \centerline{\includegraphics[width=1.0\textwidth]{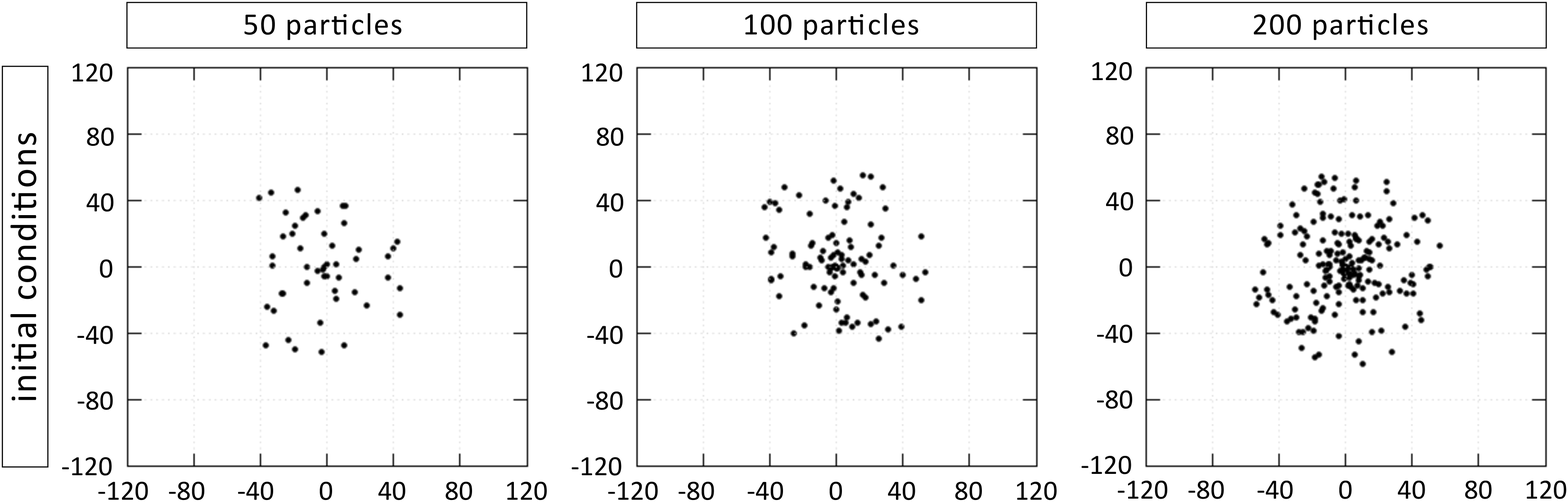}}
    \caption{
    Random spatial distributions of 50, 100 and 200 particles, respectively, used as initial conditions of the system in \eqref{eq:moto-u} to perform our second set of numerical simulations whose results are shown in Figs.~\ref{fig:6a}-\ref{fig:6d}.}
    \label{fig:5}
\end{figure}
In this perspective, the initial distributions used to perform the numerical simulation with $N = 50,\ 100,\ 200$ cells are reported in Fig.~\ref{fig:5}. It is worth to notice that, for each value of $N$, the initial distribution of the aggregate is defined regardless the value of the individual cell mass $m$. Notice also that the effect of $m$ can be absorbed in the time scale. However, we decided to keep it to show the behavior when we fix the normalization of the total mass, i.e., $M_{\textup{tot}}=m\,N$.

\begin{figure}
    \centerline{\includegraphics[width=1.0\textwidth]{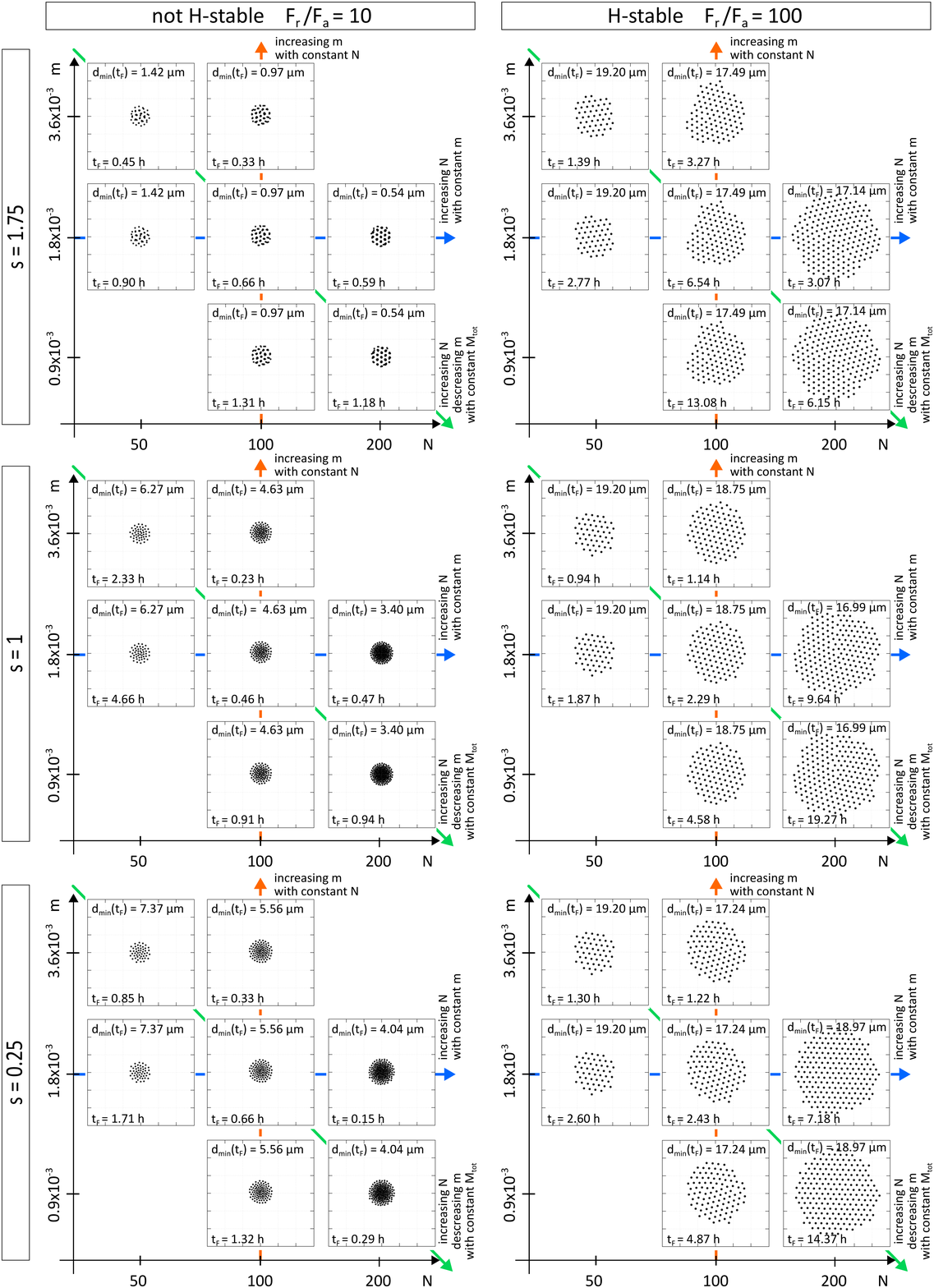}}
    \caption{
    Changes in the equilibrium configuration of the system due to the variation of the number $N$ and the individual mass $m$ of cells, both separately and simultaneously. In the former case, we consequently modify the total mass of the aggregate, while in the latter case, for any choice of the pair $\{m, N\}$ the total mass maintained equal to $0.18\unit{\mu g}$.
    Pairwise interactions are implemented by the intercellular kernel in Eq.~\eqref{eq:K-s}  with $s=1.75$ (top row), $s=1$ (central row), or $s=0.25$ (bottom row), respectively.
    The spatial distribution of the aggregate in each panel is represented within squares of size $180\times180\unit{\mu m^2}$.}
    \label{fig:6a}
\end{figure}

The numerical results obtained by setting $s=1.75$ in Eq.~\eqref{eq:K-s}, i.e., dealing with a interaction kernel that is more regular than the hyperbolic kernel at the origin, are reported in the top row in Fig.~\ref{fig:6a}. Analogously, the equilibrium configurations obtained by setting $s=1$ and $s=0.25$ in Eq.~\eqref{eq:K-s}, are reported in Fig.~\ref{fig:6a} in the central and bottom row, respectively.
The results reported in the left panels of Fig.~\ref{fig:6a} show the variation of the equilibrium configuration of the system \eqref{eq:moto-k} when the values of the interaction strengths make the interaction kernel not H-stable (i.e., $\FR/\FA > F^*$. In particular, as in Fig.~\ref{fig:4}, we consider the case with $\FR = 10\unit{\mu m/(\mu g\,s)}$ and $\FA = 1\unit{\mu m/(\mu g\,s)}$ so that $\FR/\FA = 10 < F^*$. Conversely, the results shown in the right panels of Fig.~\ref{fig:6a} refer to the case with $\FR = 100\unit{\mu m/(\mu g\,s)}$ and $\FA = 1\unit{\mu m/(\mu g\,s)}$ so that $\FR/\FA = 100 > F^*$, i.e., the interaction kernel is H-stable.

From these numerical results, it first emerges that variations of both the number of cells $N$ and the individual cell mass $m$, do not affect the characteristic pattern of the equilibrium configuration of the system in the case of both a not H-stable (clusters) and a H-stable interaction kernel (homogeneous distribution).

However, in each panel, a horizontal arrow highlights that variations in the number of component cells $N$ (with an individual fixed mass $m=0.0018\unit{\mu m/(\mu g\,s)}$) affect differently the equilibrium configuration of the system according to the H-stability of the interaction kernel. On one hand, in all the left panels of Fig.~\ref{fig:6a} (i.e., for values of the interaction strengths that makes the interaction kernel in Eq.~\eqref{eq:K-s} not H-stable), the increase of the overall number of cells $N$ forming the aggregate in fact induces a growth of both the size and the number of clusters characterizing the equilibrium configuration, whereas the equilibrium radius of the aggregate is preserved almost constant. On the other hand, in the right panel of Fig.~\ref{fig:6a} (i.e., for values of the interaction strengths that makes the interaction kernel in Eq.~\eqref{eq:K-s} H-stable), we have that the increment of the overall number of cells $N$ within the aggregate leads to a growth of the equilibrium radius of the aggregate, while the minimum intercellular relative distance is almost constant.

The effect of variations in the individual mass of cells $m$, by keeping fixed the overall number of cells $N=100$, is conversely indicated by a vertical arrow. In this case, the equilibrium configuration of the system is invariant since we can absorb $m$ in the time scale of the system, and then we only observe the right variation on the equilibration times (see the values of $t_{\textup{F}}$ reported at the bottom of each panel in Fig.~\ref{fig:6a}).

Finally, in each panel, a diagonal arrow indicates the effect of variations of both the number $N$ and the individual mass $m$ of cells constituting the aggregate, by keeping constant the total mass of the aggregate, i.e., $M_{\textup{tot}}=m\,N = 0.18\unit{\mu m}$.
In this case, the variation of the pattern of the equilibrium configuration of the system due to the increase of the number of cells $N$ (and the consequent decrease of the individual mass $m$) is consistent to the effect observed by dealing with larger aggregates of cells with the same mass $m$ (see the blue arrow). In the case of not H-stable interaction kernels, the equilibrium radius of the aggregate is almost constant in all cases, while $t_{\textup{F}}$ decreases along the green arrow. Conversely, in the case of H-stable interaction kernels, the equilibrium radius of the aggregate increases with the number of cells, while $t_{\textup{F}}$ increases along the green arrow.

\begin{figure}[ht]
    \centerline{\includegraphics[width=1.0\textwidth]{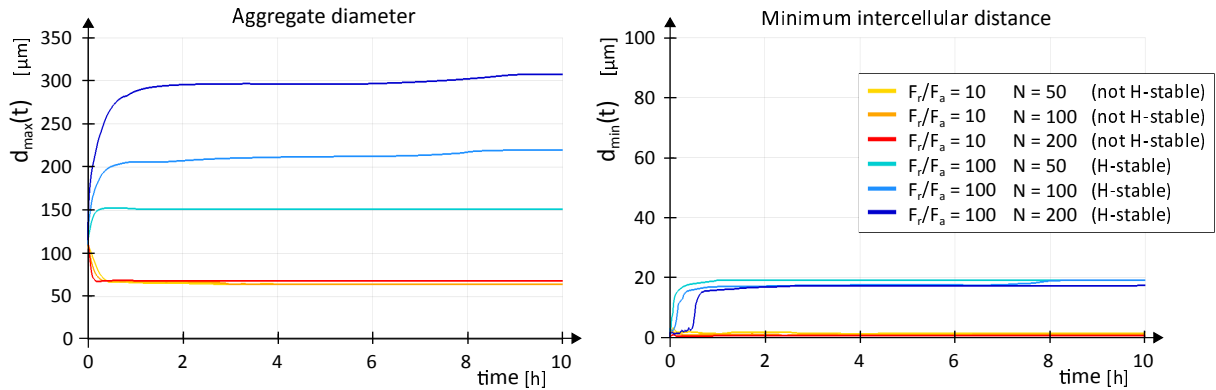}}
    \caption{
    Time evolution of the aggregate diameter $d_{\textup{max}}(t)$ and of the minimum intercellular distance $d_{\textup{min}}(t)$ arising in the case an ensemble of 50, 100, 200 particles, respectively, with individual mass $m$ fixed equal to $0.0018\unit{\mu g}$ and by setting $s=1.75$ in the interaction kernel $K$ in Eq.~\eqref{eq:K-s}.}
    \label{fig:6d}
\end{figure}

Finally, in Fig.~\ref{fig:6d}, we show the evolution in time of the minimum intercellular distance $d_{\textup{min}}(t)$ and the diameter $d_{\textup{max}}(t) = \max_{\{i,j=1,\dots,N\,:\ j\neq i\}}|\x_i(t)-\x_j(t)|$ for $s=1.75$, observed dealing with different amount of cells, i.e., $N=50,100,200$ whose the cell mass $m$ is kept equal to $0.0018\unit{\mu m}$.
As the behavior of $d_{\textup{min}}(t)$ and $d_{\textup{max}}(t)$ is similar for other values of $s$ that we do not show here for simplicity. However, as already shown in Fig.~\ref{fig:3}, the value of $s$ differently affects the minimal interparticle distance at the equilibrium $d_{\textup{min}}(t_{\textup{F}})$ according to the H-stability of the interaction kernel. Comparing the behavior of the evolutions observed for different values of $N$, we can then notice that:
\begin{itemize}
  \item[(i)] in not H-stable cases, the minimum intercellular distance at equilibrium varies with the behavior of the interaction kernel at the origin: in particular, $d_{\textup{min}}(t_{\textup{F}})$ increases as the singularity of the interaction kernel at short distances gets stronger, i.e., by decreasing the value of $s$.  However, the value of $d_{\textup{min}}(t_{\textup{F}})$ decreases to zero as $N\to\infty$ for a given $s$, see also \cite{OCBC06,CDP}. On the other hand, regardless the behavior of the interaction kernel at the origin, the equilibrium diameter of the aggregate at the equilibrium $d_{\textup{max}}(t_{\textup{F}})$ is not affected by variations in the number and the mass of cells;

\item[(ii)] conversely, in H-stable cases, the minimum intercellular distance converges to a fixed value $d_{\textup{min}}(t_{\textup{F}})$ (around $18\unit{\mu m}$ and smaller than $\dR$) depending slightly on the explicit form of the interaction kernel (i.e., the value of $s$) and the number of cells forming the aggregate $N$. Moreover, it stabilizes as $N\to \infty$ for a given $s$.
    It also follows that, regardless of the explicit form of the interaction kernel, the radius of the aggregate at the equilibrium state increases with the number of particles $N$, but it is not affected by variations of the individual mass $m$ (which only affects the equilibrium time).

\end{itemize}

%
Accounting for the above theoretical/analytical considerations, this section is devoted to a series of numerical simulations performed to show how the dynamics of pairwise interacting particles are regulated by the H-stability of the interaction kernel and its behavior at the origin. In particular, we here assume that the explicit form of the interaction kernel $K:\R_+\rightarrow\R$ in system \eqref{eq:moto-k} writes as in Eq.~\eqref{eq:K-s} and perform several numerical tests by varying either the value of $s$ (that characterizes the repulsive behavior of $K$ at short distances) or the ratio $\FR/\FA$ between the repulsive and adhesive interaction strength (that conversely defines the H-stability of the interaction kernel).

\bigskip
\noindent
In this perspective, let us first consider a cell aggregate constituted by $N=100$ individuals. In all realizations, cells are initially randomly distributed within a round area of radius equal to $40\unit{ \mu m}$ and centered at the origin, as shown in Fig.~\ref{fig:3} (left panel). In particular, the initial distribution of the aggregate $\X(0)$ is such that for each cell $i=1,\dots,N$ the minimal interparticle distance $d_{\textup{min}}(0)$, defined in Eq.~\eqref{eq:dmin}, is not null (in order to avoid that the center of mass of distinct cells are initially located in the same position) and lower than $\dA$ (so that each cell is initially able to interact at least with another cell). According to \cite{CaSmPl_JMB2017} and the references therein, we hereafter set the cell biophysical properties (mass and size) as follows:
    \begin{center}
    \begin{tabular}{clcc}
      Param. & Description & Value [Unit] & Ref.\\
      \hline
      \noalign{\smallskip}
        $m$ & mean cell mass & $1.8\cdot10^{-3}\unit{ \mu g}$
                & \cite{CaSmPl_MMNP2015}\\
        $d_{\textup{r}}$ & mean cell diameter & $20\unit{ \mu m}$
                & \cite{CaSmPl_MMNP2015}\\
        $d_{\textup{a}}$ & maximal extension of cell filopodia & $60\unit{ \mu m}$
                & \cite{CaSmPl_MMNP2015}\\
      \hline
      \end{tabular}
    \end{center}
\begin{figure}[ht]
    \centerline{\includegraphics[width=1.0\textwidth]{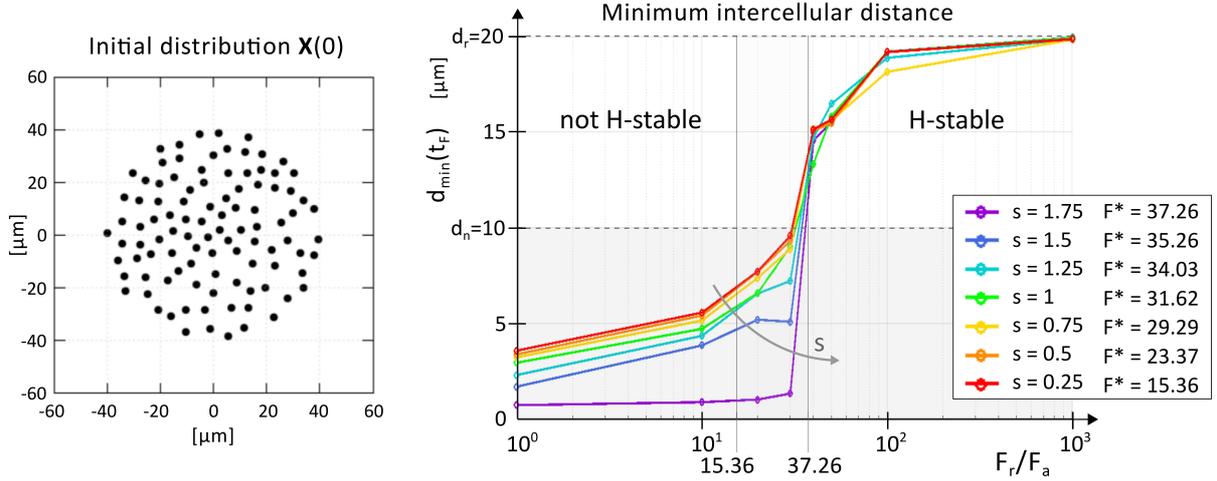}}
    \caption{
    Left panel: Representation of the random spatial distribution of 100 individuals that constitutes the initial condition $\X(0)$ in all realizations of our first series of numerical realizations.
    As a remark, in all simulations, particles are able to move over all the planar space $\R^2$ and no boundary constraints are required.
    Right panel: Variation of the minimum intercellular distance at the equilibrium, i.e., $d_{\textup{min}}(t_{\textup{F}})$ at the equilibrium (see Eq.~\eqref{eq:dmin}), according to different values $\FR/\FA$, observed for distinct choices of $s$, i.e., of the behavior of the pairwise interaction kernel/potential at the origin. 
    According to Theorem~\ref{thm-loc}, the interaction kernel/potential is not H-stable when $\FR/\FA<F^*$ and H-stable when $\FR/\FA>F^*$. For each choice of $s$, the threshold value $F^*$ is reported in the legend.
    The shaded region finally represents unsustainable stationary intercellular distances, i.e., that result in too much packed component cells.
    }
    \label{fig:3}
\end{figure}
From the analytical results provided in the previous section, it thereby follows that the H-stability of the interaction kernel $K$ depends on the ratio of the interaction strengths $\FR$ and $\FA$, according to the value of $s$, i.e., to the shape of the repulsive part of the interaction kernel. Specifically, we here investigate the following cases:
\begin{center}
    \begin{tabular}{c|cccccccc}
        \noalign{\smallskip}
        \hline
        $s$   & 2      & 1.75 & 1.5 & 1.25 & 1 & 0.75 & 0.5 & 0.25 \\
        $F^*$ & 38.40  & 37.26 & 35.26 & 34.02 & 31.32 & 28.29 & 23.37 & 15.36 \\
        \hline
  \end{tabular}
 \end{center}
where $F^*$ is the minimum ratio between the repulsive and adhesive interaction strengths that makes the interaction kernel H-stable (see Eq.~\eqref{eq:F*s} for cases with $s\neq1$ and Eq.~\eqref{eq:F*1} when $s=1$).
In this respect, we first perform a series of numerical simulations exploring the effect of variations of $\FR/\FA$ for each one of the above-cited values of $s$. This allows us to observe how the equilibrium configuration of the system and cell dynamics is regulated by the H-stability of a given interaction kernel/potential.
In more details, in all realizations, we keep the strength of cell-cell adhesiveness $\FA$ constantly equal to $1\unit{ \mu m/(\mu g\,s)}$, so that the value of cell resistance to compression $\FR$ coincides with $\FR/\FA$ and determines the H-stability of $K$. This means that we are fixing the adhesive characteristics of cell-cell interactions, focusing on the repulsive part. However, the discriminating parameter is still the ratio $\FR/\FA$.
In particular, we test all the following values for the repulsive strength: $\FR = 1,\ 10,\ 20,\ 30,\ 40,\ 50,\ 100,\ 1000 \unit{ \mu m/(\mu g s)}$. For all realizations, the evolution of the system has been observed until a stable equilibrium configuration is established and $d_{\textup{min}}(t_F)$ denotes the minimal interparticle distance evaluated at the equilibrium.
The graph in the right panel of Fig.~\ref{fig:3} then shows how the final minimal intercellular distance $d_{\textup{min}}(t_F)$ varies according to the value of the ratio $\FR/\FA$ for each one of considered value of $s$.
According to the above theoretical considerations, it clearly emerges that, independently on the slope of the repulsive part of the kernel (i.e., on $s$), the H-stability of the interaction kernel $K$ regulates the minimum distance between the interacting particles at the equilibrium configuration.
In fact, for any choices of the shape of $K$, we have that:
if $\FR/\FA < F^*$ (i.e., $K$ is not-H stable), the minimal intercellular distance at the equilibrium $d_{\textup{min}}(t_{\textup{F}})$ falls within the range $[0, \dN]$;
while, if $\FR/\FA > F^*$ (i.e., $K$ is H stable), $d_{\textup{min}}(t_{\textup{F}})\in[\dN, \dR]$.
From a biological point of view, these results can be interpreted as it follows:
\begin{itemize}
  \item[-] \emph{not H-stable interaction kernels} (i.e., $\FR/\FA < F^*$) are not able to avoid the superposition of cell nuclei, and thereby the observed equilibrium configurations are biologically unreliable (i.e., $d_{\textup{min}}<\dN$);
  \item[-] conversely, \emph{H-stable interaction kernels} (i.e., $\FR/\FA > F^*$) allow cells to preserve a equilibrium intercellular distance large enough to survive (i.e., $d_{\textup{min}}>\dN$).
\end{itemize}

In this respect, we can further observe that, starting from an initial distribution of cells constituting a single cluster, the minimal intercellular distance at equilibrium will never exceed the repulsion radius $\dR$, i.e., $d_{\textup{min}}(t_{\textup{F}})\leq\dR$, regardless of both the shape and the H-stability of the interaction kernel. In fact, dealing with a first order model where cell dynamics is regulated solely by repulsive-attractive interactions, we have that as $d_{\textup{min}}(t)$ exceeds $\dR$, the repulsive velocity component between any pair of cells vanishes and cells not more drift away.

Referring again to the graph in the right panel of Fig.~\ref{fig:3}, we can moreover notice that in case of not H-stable interaction kernels/potentials the minimal intercellular distance at the equilibrium varies according to the singularity of the repulsive part of $K$. In fact, comparing the results obtained by setting the interaction strengths such that $\FR/\FA < 15.36$ (so that the interaction kernels/potentials are not H-stable for all considered choices of the value of $s$), we have that the value of $d_{\textup{min}}(t_{\textup{F}})$ increases as $s$ decreases, i.e., as $K$ is more singular near to the origin.
On the other hand, we do not observe an analogous correlation between the minimal interparticle distance at the equilibrium and the singularity of the kernel/potential near to the origin in case of H-stable interaction kernels, i.e., when the interaction strength are such that $\FR/\FA > 37.26$ (see again Fig.~\ref{fig:3}, right panel).

\begin{figure}[ht]
    \centerline{\includegraphics[width=1.0\textwidth]{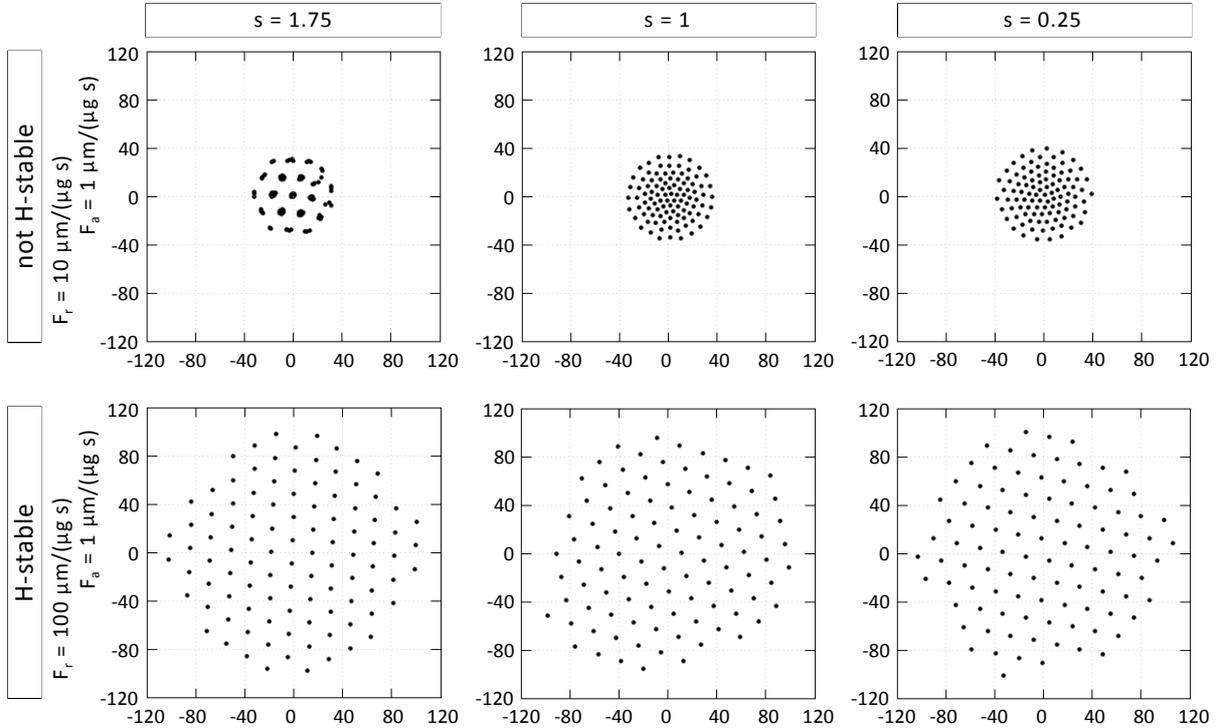}}
    \caption{
    Equilibrium configuration of the cell aggregate (i.e., at $t_F = 8\unit{h}$) observed in the case of some representative choices of both the shape of the interaction kernel at the origin (which is characterized by the value of $s$, i.e., $s = 1.75,\ 1,\ 0.25$), and the ratio between the interaction strengths parameter values (i.e., $\FR/\FA$, which makes the interaction kernels H-stable or not). Specifically, the adhesive strength $\FA$ is here constantly maintained equal to $1\unit{\mu m/(\mu g\,s)}$, so that the value of the ratio $\FR/\FA$ coincides with the value of the repulsive strength $\FR$.
    }
    \label{fig:4}
\end{figure}

In order to further highlight the above commented difference between the not H-stable and H-stable regimes, we report in Fig.~\ref{fig:4} the equilibrium configuration of the aggregate observed in the some representative cases among those described above. Specifically, we hereafter focus only on three possible choices for the behavior of the interaction kernel near the origin (i.e., referring to Eq.~\eqref{eq:K-s}, those corresponding to $s=1.75,\ 1,\ 0.25$), and only to two values of the ratio $\FR/\FA$, i.e., $\FR/\FA = 10$ and $100$, that respectively makes the considered interaction kernels not H-stable and H-stable.
Let us moreover notice that we are comparing the results arising when the interaction kernel at the origin is assumed either more regular than hyperbolic ($s=1.75$), hyperbolic ($s=1$) or more singular than hyperbolic ($s=0.25$).
These final distributions further show that if the interaction kernel is not H-stable (as in the first row in Fig.~\ref{fig:4}), the individuals tend to clusterize and the final minimal intercellular distance depends on the behavior of the interaction kernel at the origin, i.e., on $s$. In particular, we have that by setting $s=1.75$, cells divide into several clusters, with very small intercellular distances within the clusters. Further, the spatial distribution of cells at the equilibrium appears more homogeneous as the value of $s$ is lower, see the cases with $s=1$ and $s=0.25$, i.e., as the interaction kernel/potential is more singular at the origin. However, in all cases, the intercellular distance is always too small to allow cell survival.
On the other hand, dealing with a H-stable interaction kernel (see the second row in Fig.~\ref{fig:4}), cells re-organize in a quite homogeneous distribution characterized by a minimum intercellular distance that allows cell survival. Moreover, such behavior arises for all values of the parameter $s$ here reported as well as those considered above (i.e., $s=0.5,\,0.75,\,1.25\,1.5$).

\bigskip

To investigate deeper how the choice of a not H-stable interaction kernel rather than a H-stable one affects the system dynamics, we focus on the six representative cases reported in Fig.~\ref{fig:4} and analyze the effect of variations of the overall number $N$ and of the individual mass $m$ of cells, both separately (thereby varying the overall mass of the aggregate) and simultaneously by keeping constant the overall mass of the aggregate $M_{\textup{tot}}=mN$.
In more details, for each one of the representative cases considered in Fig.~\ref{fig:4}, we perform a series of numerical simulations to compare the above-commented equilibrium configurations (obtained with $N=100$ cells with mass $m=0.0018 \unit{\mu g}$) with those arising if we deal with:
\begin{itemize}
  \item[(i)] a smaller/larger aggregate of cells whose individual mass $m$ is still equal to $0.0018 \unit{\mu g}$ (specifically, we investigate the cases with $N=50$ and $N=200$ cells);
  \item[(ii)] an aggregate still formed by $N=100$ cells whose individual mass $m$ is lower/higher than $0.0018 \unit{\mu g}$ (in particular, we consider cells with $m=0.0009 \unit{\mu g}$ and $m=0.0036 \unit{\mu g}$, respectively);
  \item[(iii)] an aggregate whose overall mass $M_{\textup{tot}}=m\,N$ is kept equal to $0.18 \unit{\mu g}$, but the number $N$ and, in turn, the individual mass $m$ of cells are different from $100$ and $0.0018 \unit{\mu g}$, respectively, (in this case, we first consider an aggregate composed by $N=50$ cells with $m=0.0036\unit{\mu g}$; then we set $N=200$ and $m=0.0009\unit{\mu g}$). This is related to the mean-field limit as explained in Section~\ref{sec:2}.
\end{itemize}
\begin{figure}[ht]
    \centerline{\includegraphics[width=1.0\textwidth]{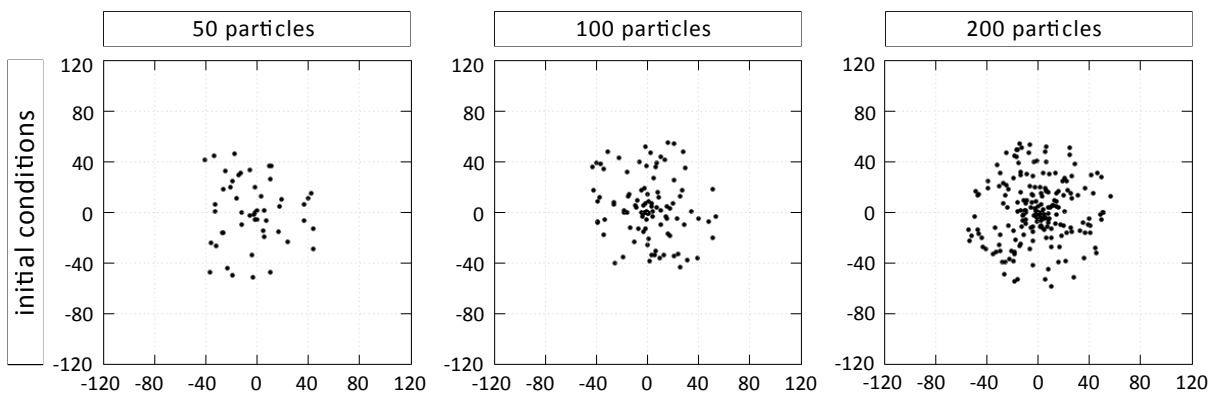}}
    \caption{
    Random spatial distributions of 50, 100 and 200 particles, respectively, used as initial conditions of the system in \eqref{eq:moto-u} to perform our second set of numerical simulations whose results are shown in Figs.~\ref{fig:6a}-\ref{fig:6d}.}
    \label{fig:5}
\end{figure}
In this perspective, the initial distributions used to perform the numerical simulation with $N = 50,\ 100,\ 200$ cells are reported in Fig.~\ref{fig:5}. It is worth to notice that, for each value of $N$, the initial distribution of the aggregate is defined regardless the value of the individual cell mass $m$. Notice also that the effect of $m$ can be absorbed in the time scale. However, we decided to keep it to show the behavior when we fix the normalization of the total mass, i.e., $M_{\textup{tot}}=m\,N$.

\begin{figure}
    \centerline{\includegraphics[width=1.0\textwidth]{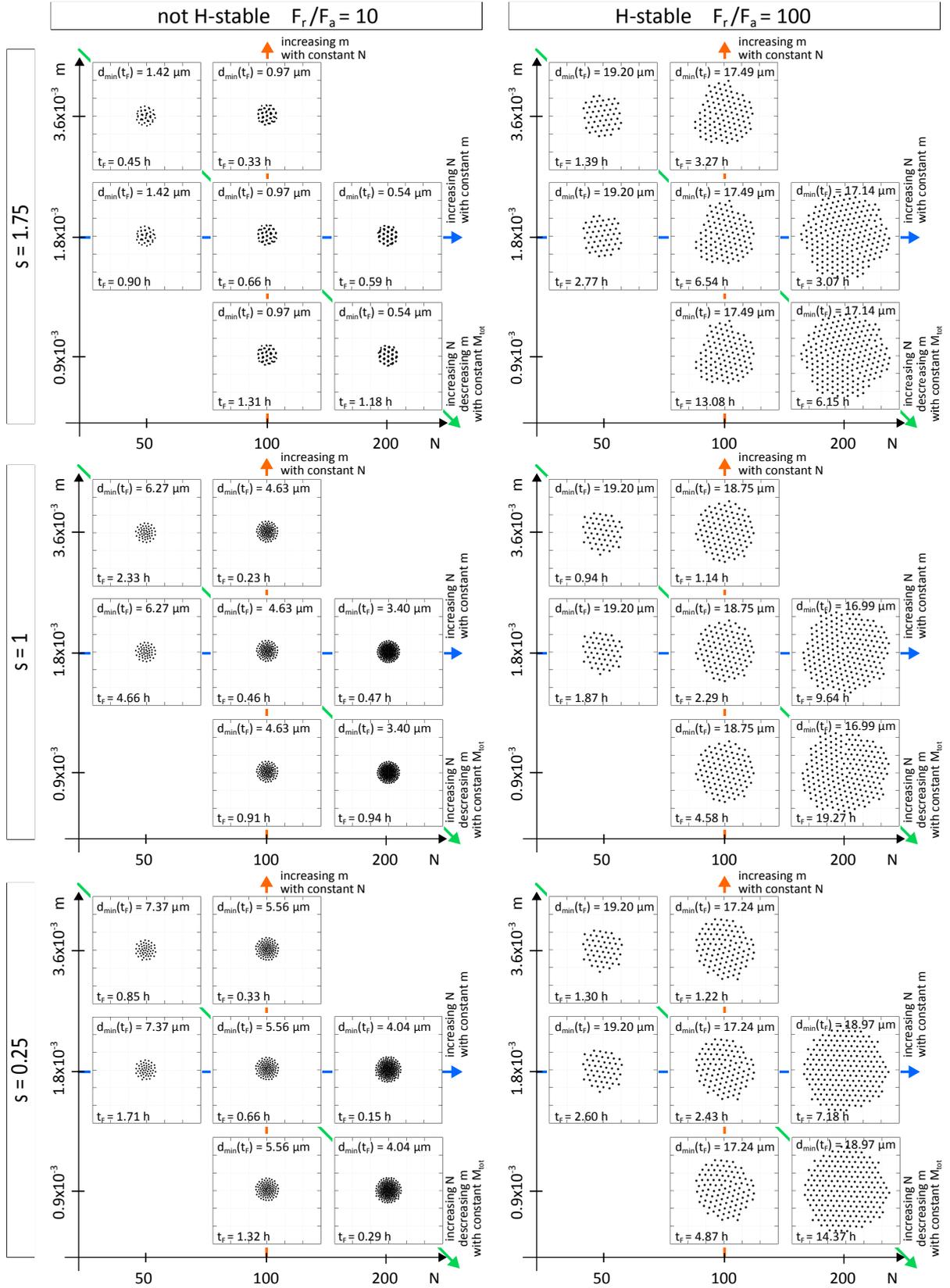}}
    \caption{
    Changes in the equilibrium configuration of the system due to the variation of the number $N$ and the individual mass $m$ of cells, both separately and simultaneously. In the former case, we consequently modify the total mass of the aggregate, while in the latter case, for any choice of the pair $\{m, N\}$ the total mass maintained equal to $0.18\unit{\mu g}$.
    Pairwise interactions are implemented by the intercellular kernel in Eq.~\eqref{eq:K-s}  with $s=1.75$ (top row), $s=1$ (central row), or $s=0.25$ (bottom row), respectively.
    The spatial distribution of the aggregate in each panel is represented within squares of size $180\times180\unit{\mu m^2}$.}
    \label{fig:6a}
\end{figure}

The numerical results obtained by setting $s=1.75$ in Eq.~\eqref{eq:K-s}, i.e., dealing with a interaction kernel that is more regular than the hyperbolic kernel at the origin, are reported in the top row in Fig.~\ref{fig:6a}. Analogously, the equilibrium configurations obtained by setting $s=1$ and $s=0.25$ in Eq.~\eqref{eq:K-s}, are reported in Fig.~\ref{fig:6a} in the central and bottom row, respectively.
The results reported in the left panels of Fig.~\ref{fig:6a} show the variation of the equilibrium configuration of the system \eqref{eq:moto-k} when the values of the interaction strengths make the interaction kernel not H-stable (i.e., $\FR/\FA > F^*$. In particular, as in Fig.~\ref{fig:4}, we consider the case with $\FR = 10\unit{\mu m/(\mu g\,s)}$ and $\FA = 1\unit{\mu m/(\mu g\,s)}$ so that $\FR/\FA = 10 < F^*$. Conversely, the results shown in the right panels of Fig.~\ref{fig:6a} refer to the case with $\FR = 100\unit{\mu m/(\mu g\,s)}$ and $\FA = 1\unit{\mu m/(\mu g\,s)}$ so that $\FR/\FA = 100 > F^*$, i.e., the interaction kernel is H-stable.

From these numerical results, it first emerges that variations of both the number of cells $N$ and the individual cell mass $m$, do not affect the characteristic pattern of the equilibrium configuration of the system in the case of both a not H-stable (clusters) and a H-stable interaction kernel (homogeneous distribution).

However, in each panel, a horizontal arrow highlights that variations in the number of component cells $N$ (with an individual fixed mass $m=0.0018\unit{\mu m/(\mu g\,s)}$) affect differently the equilibrium configuration of the system according to the H-stability of the interaction kernel. On one hand, in all the left panels of Fig.~\ref{fig:6a} (i.e., for values of the interaction strengths that makes the interaction kernel in Eq.~\eqref{eq:K-s} not H-stable), the increase of the overall number of cells $N$ forming the aggregate in fact induces a growth of both the size and the number of clusters characterizing the equilibrium configuration, whereas the equilibrium radius of the aggregate is preserved almost constant. On the other hand, in the right panel of Fig.~\ref{fig:6a} (i.e., for values of the interaction strengths that makes the interaction kernel in Eq.~\eqref{eq:K-s} H-stable), we have that the increment of the overall number of cells $N$ within the aggregate leads to a growth of the equilibrium radius of the aggregate, while the minimum intercellular relative distance is almost constant.

The effect of variations in the individual mass of cells $m$, by keeping fixed the overall number of cells $N=100$, is conversely indicated by a vertical arrow. In this case, the equilibrium configuration of the system is invariant since we can absorb $m$ in the time scale of the system, and then we only observe the right variation on the equilibration times (see the values of $t_{\textup{F}}$ reported at the bottom of each panel in Fig.~\ref{fig:6a}).

Finally, in each panel, a diagonal arrow indicates the effect of variations of both the number $N$ and the individual mass $m$ of cells constituting the aggregate, by keeping constant the total mass of the aggregate, i.e., $M_{\textup{tot}}=m\,N = 0.18\unit{\mu m}$.
In this case, the variation of the pattern of the equilibrium configuration of the system due to the increase of the number of cells $N$ (and the consequent decrease of the individual mass $m$) is consistent to the effect observed by dealing with larger aggregates of cells with the same mass $m$ (see the blue arrow). In the case of not H-stable interaction kernels, the equilibrium radius of the aggregate is almost constant in all cases, while $t_{\textup{F}}$ decreases along the green arrow. Conversely, in the case of H-stable interaction kernels, the equilibrium radius of the aggregate increases with the number of cells, while $t_{\textup{F}}$ increases along the green arrow.

\begin{figure}[ht]
    \centerline{\includegraphics[width=1.0\textwidth]{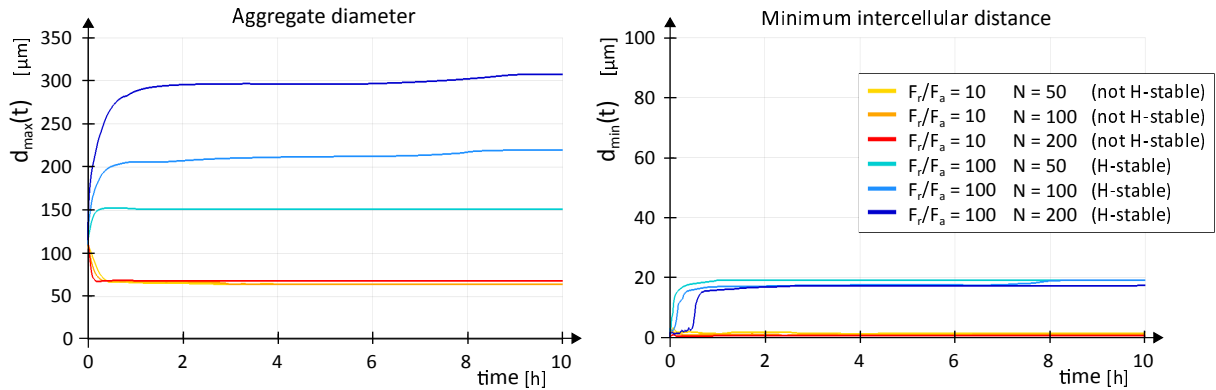}}
    \caption{
    Time evolution of the aggregate diameter $d_{\textup{max}}(t)$ and of the minimum intercellular distance $d_{\textup{min}}(t)$ arising in the case an ensemble of 50, 100, 200 particles, respectively, with individual mass $m$ fixed equal to $0.0018\unit{\mu g}$ and by setting $s=1.75$ in the interaction kernel $K$ in Eq.~\eqref{eq:K-s}.}
    \label{fig:6d}
\end{figure}

Finally, in Fig.~\ref{fig:6d}, we show the evolution in time of the minimum intercellular distance $d_{\textup{min}}(t)$ and the diameter $d_{\textup{max}}(t) = \max_{\{i,j=1,\dots,N\,:\ j\neq i\}}|\x_i(t)-\x_j(t)|$ for $s=1.75$, observed dealing with different amount of cells, i.e., $N=50,100,200$ whose the cell mass $m$ is kept equal to $0.0018\unit{\mu m}$.
As the behavior of $d_{\textup{min}}(t)$ and $d_{\textup{max}}(t)$ is similar for other values of $s$ that we do not show here for simplicity. However, as already shown in Fig.~\ref{fig:3}, the value of $s$ differently affects the minimal interparticle distance at the equilibrium $d_{\textup{min}}(t_{\textup{F}})$ according to the H-stability of the interaction kernel. Comparing the behavior of the evolutions observed for different values of $N$, we can then notice that:
\begin{itemize}
  \item[(i)] in not H-stable cases, the minimum intercellular distance at equilibrium varies with the behavior of the interaction kernel at the origin: in particular, $d_{\textup{min}}(t_{\textup{F}})$ increases as the singularity of the interaction kernel at short distances gets stronger, i.e., by decreasing the value of $s$.  However, the value of $d_{\textup{min}}(t_{\textup{F}})$ decreases to zero as $N\to\infty$ for a given $s$, see also \cite{OCBC06,CDP}. On the other hand, regardless the behavior of the interaction kernel at the origin, the equilibrium diameter of the aggregate at the equilibrium $d_{\textup{max}}(t_{\textup{F}})$ is not affected by variations in the number and the mass of cells;

\item[(ii)] conversely, in H-stable cases, the minimum intercellular distance converges to a fixed value $d_{\textup{min}}(t_{\textup{F}})$ (around $18\unit{\mu m}$ and smaller than $\dR$) depending slightly on the explicit form of the interaction kernel (i.e., the value of $s$) and the number of cells forming the aggregate $N$. Moreover, it stabilizes as $N\to \infty$ for a given $s$.
    It also follows that, regardless of the explicit form of the interaction kernel, the radius of the aggregate at the equilibrium state increases with the number of particles $N$, but it is not affected by variations of the individual mass $m$ (which only affects the equilibrium time).
\end{itemize}

\section{Biological applications}
\label{sec:4}

\subsection{Inverse problem}

\begin{figure}
    \centerline{\includegraphics[width=1.0\textwidth]{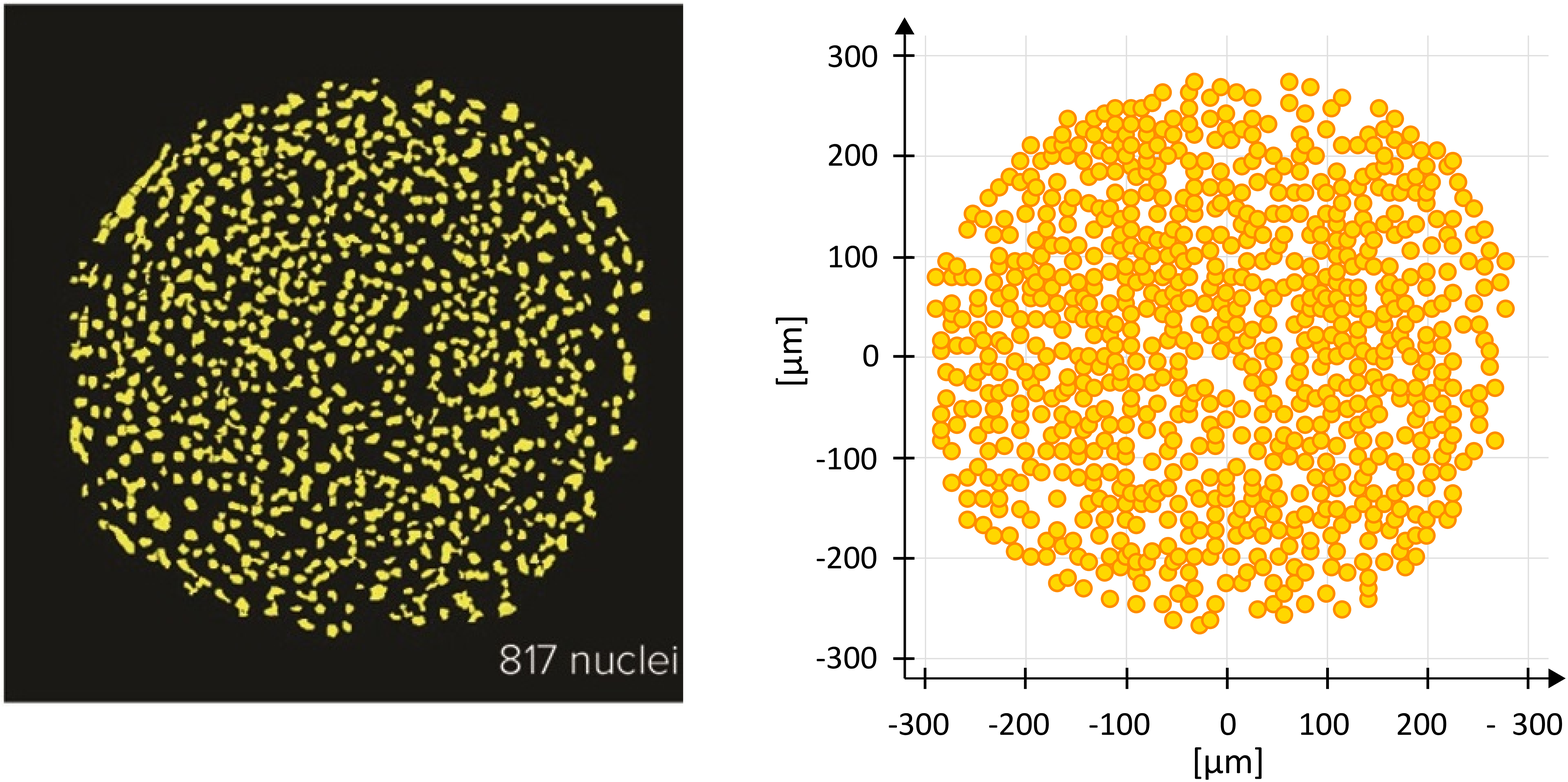}}
    \caption{
    Left panel:
    Round colony of ovarian cancer cells (OVCAR:39) plated on a two dimensional Petri dish in physiological liquid. Experimental image kindly provided by Prof. Luca Munaron of the Department of Life Sciences and Systems Biology, Universit\`a degli Studi di Torino (Italy). It is a frame of an experimental assay where a spheroid of ovarian cancer cells is plated on a two dimensional Petri dish.
    Right panel: Representation of the spatial distribution $\textbf{X}_{\textup{R}}$ of cells given by the experiment.
    }
    \label{fig:7}
\end{figure}

Accounting for the above-considerations and numerical results, this section focuses on how to find a proper interaction kernel so that the typical equilibrium state of system in \eqref{eq:moto-k} resembles a cell distribution observed in a given experiments. In particular, we here consider the biological picture shown in Fig.~\ref{fig:7}: it is a frame of an experimental assay where a spheroid of ovarian cancer cells is placed on a two dimensional Petri dish (kindly provided by kindly provided by Prof. Luca Munaron of the Department of Life Sciences and Systems Biology, Universit\`a degli Studi di Torino, Italy). This picture gives the spatial distribution of the nuclei (yellow dots) of $N=817$ cells constituting the spheroid, and the aim of this section is to identify (at least) one of the interaction kernels defined in Eq.~\eqref{eq:K-s}, as well as the values of the interaction parameters that is able to give a good approximation of the experimental distribution of cells shown in Fig.~\ref{fig:7}.
In this perspective, by assuming that each material point represents the center of mass of a single cell, it is first worth to notice that the proper interaction kernel has to ensure that the minimum intercellular distance required for cell survival is preserved during all system evolution. In this respect, the numerical results shown in previous section suggest that it is preferable to opt for a H-stable interaction kernel rather than for a not H-stable one. We therefore perform a series of simulations by assuming H-stable interaction kernels to implement intercellular interactions and compare the resulting equilibrium state of a system of 817 cells with the experimental cell distribution in Fig.~\ref{fig:7}.

\begin{figure}[ht]
    \centerline{\includegraphics[width=1.0\textwidth]{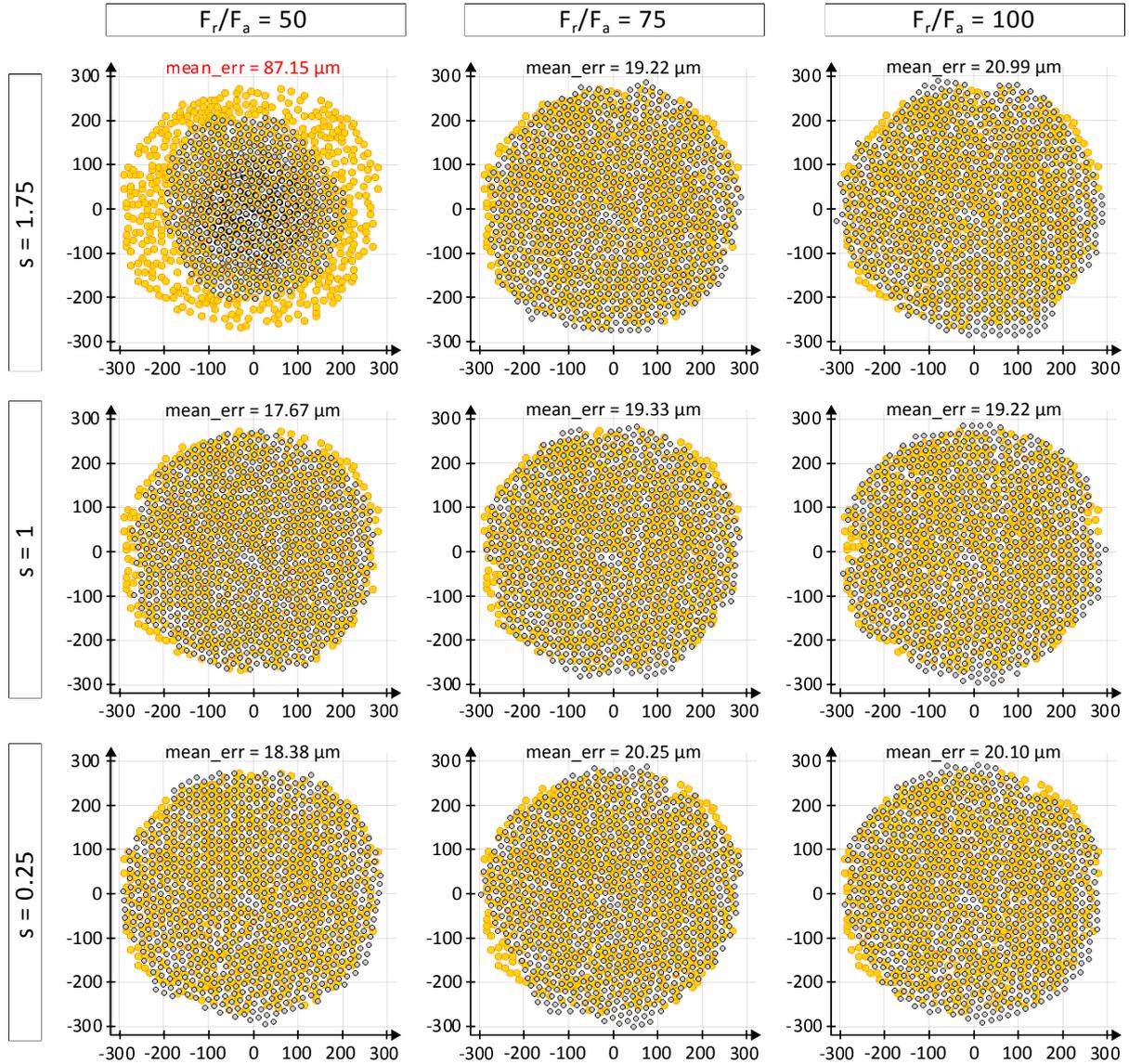}}
    \caption{
    Comparison of the spatial distribution of cells observed in the experiment (yellow circles) and the spatial distribution at the equilibrium obtained by the numerical simulations. In each case, the value of $\textup{mean}\_\textup{err}$ quantifies the mean distance between the numerical and the experimental distribution of particles.
    }
    \label{fig:8}
\end{figure}

Entering in more details, we here focus again only on three possible choices for the explicit form of the interaction kernel at the origin, i.e., referring to Eq.~\eqref{eq:K-s}, with those denoted by $s = 1.75,\ 1,\ 0.25$, respectively, and for each of these cases we take into account three distinct settings of the interaction parameters such that $\FR/\FA > F^*$ ensuring the H-stability of the interaction kernel.
In particular, all numerical simulations are performed assuming $\dR=20\unit{\mu m}$, $\dA=60\unit{\mu m}$ and $\FA=1\unit{\mu m /(\mu g\,s)}$, as in previous section, whereas the repulsive strength $\FR$ is respectively set equal to $50,\ 75, \ 100\unit{\mu m /(\mu g\,s)}$.
Finally, the mass $m$ of cancer cell is fixed equal to $0.0018\unit{\mu m/(\mu g\,s)}$, according again to \cite{CaSmPl_JMB2017}, and all numerical simulations are initialized from the same initial distribution of cells.
The equilibrium distributions obtained in these cases are shown in Fig.~\ref{fig:8}. Each panel refers to a different choice of both the behavior of the interaction kernel at the origin (i.e., the value of $s$) and of the repulsive strength $\FR$, and for each case we plot together both the real position of cells (as represented in Fig.~\ref{fig:7}) and the equilibrium distribution given by the numerical simulations.

It clearly emerges that the worst approximation of the real distribution of cells occurs by setting $s=1.75$ and $\FR/\FA=50$, (i.e., top left panel in Fig.~\ref{fig:8}), while it results not obvious how to chose between the other cases. In this respect, in order to compare the real distribution and the numerical equilibrium state, we use the Hungarian algorithm (also known as Munkres assignment algorithm), which is a combinational optimization algorithm that return a value $C$, termed minimum cost, such that $\textup{mean}\_\textup{err} = C/817$ quantifies the mean distance between the numerical and the real position of each cell. We then report the values of the mean error $\textup{mean}\_\textup{err}$, at the top of each panel in Fig.~\ref{fig:8}. Consistently with our above considerations about the difference between the numerical results and the real distribution of cells, the higher value of  $\textup{mean}\_\textup{err}$ occurs in the case characterized by $s=1.75$ and $\FR/\FA=50\unit{\mu m/(\mu g\,s)}$ (see again the top left panel in Fig.~\ref{fig:8}), whereas in all other cases the values of $\textup{mean}\_\textup{err}$ are similar, and moreover the values of $\textup{mean}\_\textup{err}$ is always close (and also lower than) the value of $\dR$.
From these results, we can conclude that, dealing with these possible options to implement the interaction cells, the better approximation of the real cell distribution of cells occurs by assuming that the interaction kernel is hyperbolic at the origin (i.e., $s=1$ in Eq.~\eqref{eq:K-s}) and by setting the repulsive strength $\FR/\FA$ equal to $50\unit{\mu m/(\mu g\,s)}$ (see the bottom left panel in Fig.~\ref{fig:8}). However, it is worth to notice that we here compare few possible options for the explicit form of the interaction kernel and, moreover, we do not take into account the dependence of the final configuration on the initial condition used in numerical simulations. It is therefore possible to improve the approximation of the experimental distribution of cells by replicating the above method for different assumptions about the behavior of $K$ and taking into account the dynamics by also comparing the distribution of cells in time. This is just a first step in the direction of identifying interaction potentials from experimental data according to the specific problem we are dealing with.

\subsection{Cell sorting}

In multicellular organisms, the relative adhesion of various cell types to each other or to noncellular components surrounding them is also fundamental. From the late 1950s, it has been widely noticed that during embryonic development the behavior of cell aggregates resembles that of viscous fluid. A random mixture of two types of embryonic cells, in fact, spontaneously reorganizes to reestablish coherent homogeneous tissues. A similar process is a key step also in the regeneration of normal animal from aggregates of dissociated cells of adult hydra. It also explains the layered structure of the embryonic retina. These phenomena, commonly called \emph{cell sorting}, involve neither cell division nor differentiation, but are entirely caused by spatial rearrangements of cell positions due to differences in the specific adhesivities; see \cite{CPM_GfGa_PRL1992} and references therein. Indeed, specific hierarchies of adhesive strengths lead to specific configurations of the cellular aggregate.

\begin{figure}[ht]
    \centerline{\includegraphics[width=0.4\textwidth]{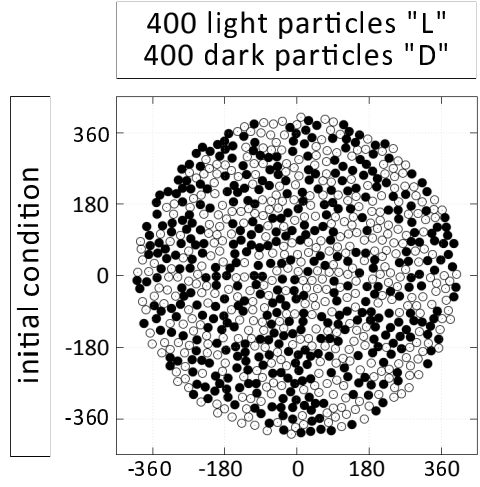}}
    \caption{
    Initial distribution of the representative two- population aggregate (i.e., composed of light ``L" and dark ``D" cells). The aggregate is constituted by 800 cells divided in two groups equal in number. The initial condition has been randomly generated such that the minimal intercellular distance at the initial instant is greater than $\dR$ (i.e., cell overlapping is initially avoided).
    }
    \label{fig:9}
\end{figure}

\begin{figure}
    \centerline{\includegraphics[width=1.0\textwidth]{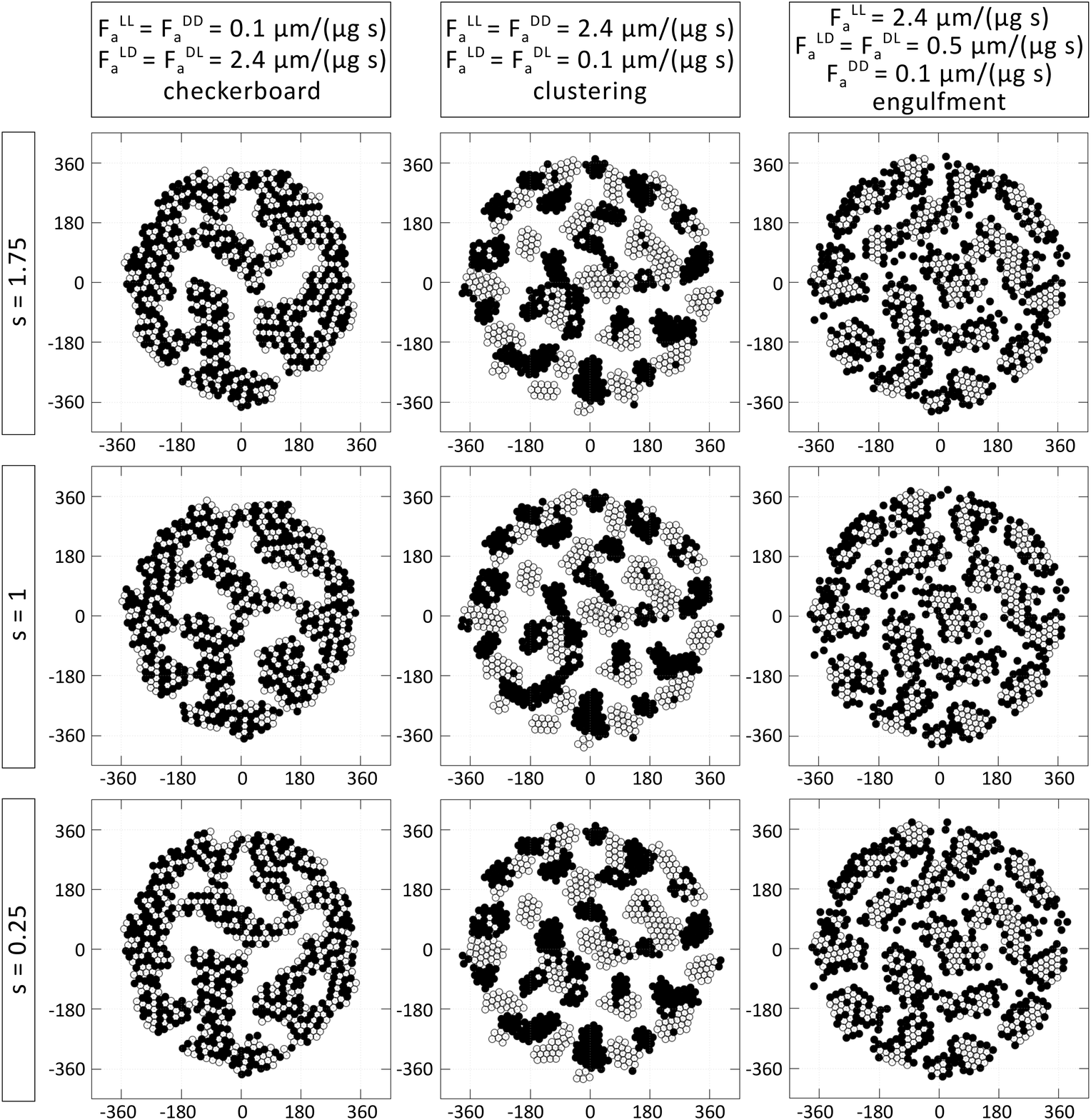}}
    \caption{
    Final distribution of the two-population aggregate (i.e., composed of light ``L" and dark ``D" cells) for distinct choices of both the behavior of the interaction kernel at the origin, i.e., the value of $s$ in Eq.~\eqref{eq:K-s} and the values of adhesion strengths. The choice of $s$ does not affect the final configuration as long as the H-stability is satisfied.
    }
    \label{fig:10}
\end{figure}

\noindent
A simple and intuitive simulation reproducing biological cell sorting consists of a cellular aggregate formed by two types of randomly positioned individuals, namely, light ``L" and dark ``D" (that are graphically represented by white and black circles, respectively, see Fig.~\ref{fig:9}).
In particular, we here assume that the two types of cells have the same biophysical properties (i.e., cell mass $m=0.0018\unit{\mu g}$, cell diameter $d_R=20\unit{\mu m}$ and maximal extension of filopodia $d_A=60\unit{\mu m}$) but that behave differently. In this respect, the evolution in time of the spatial distribution of the aggregate is given by the extension of the system in Eq.~\eqref{eq:moto-k} to two populations:
\begin{equation*}
      \begin{array}{l}
      \ds\dfrac{d\x_i^{\textup{L}}}{dt} = -
            m\sum_{j=1\atop j\neq i}^{N_\textup{L}}\, K^{\textup{LL}}(|\x_i^{\textup{L}}(t) - \x_j^{\textup{L}}(t)|)\dfrac{\x_i^{\textup{L}}(t) - \x_j^{\textup{L}}(t)}{|\x_i^{\textup{L}}(t) - \x_j^{\textup{L}}(t)|}- m\sum_{j=1\atop }^{N_\textup{D}}\, K^{\textup{LD}}(|\x_i^{\textup{D}}(t) - \x_j^{\textup{L}}(t)|)\dfrac{\x_i^{\textup{D}}(t) - \x_j^{\textup{L}}(t)}{|\x_i^{\textup{D}}(t) - \x_j^{\textup{L}}(t)|};\\[7mm]
      \ds\dfrac{d\x_h^{\textup{D}}}{dt} =
            -m\sum_{j=1\atop j\neq h}^{N_\textup{D}}\, K^{\textup{DD}}(|\x_h^{\textup{D}}(t) - \x_j^{\textup{D}}(t)|)\dfrac{\x_h^{\textup{D}}(t) - \x_j^{\textup{D}}(t)}{|\x_h^{\textup{D}}(t) - \x_j^{\textup{D}}(t)|} - m\sum_{j=1\atop }^{N_\textup{L}}\, K^{\textup{DL}}(|\x_h^{\textup{L}}(t) - \x_j^{\textup{D}}(t)|)\dfrac{\x_h^{\textup{L}}(t) - \x_j^{\textup{D}}(t)}{|\x_h^{\textup{L}}(t) - \x_j^{\textup{D}}(t)|},
      \end{array}
\end{equation*}
where $\x_i^{\textup{L}}(t)$ and $\x_h^{\textup{D}}(t)$, with $i=1,\dots,N_{\textup{L}}$ and $h=1,\dots,N_{\textup{D}}$ denote the actual positions of the ``L" and ``D" cells, respectively; while the interaction kernels $K^{pq}:\R_+\mapsto\R$, with $p,q\in\{$L,D$\}$, denote how a cell of phenotype $p$ reacts to the presence of a cell of phenotype $q$.
Specifically, we assume that both ``L" and ``D" cells are characterized by the same resistance to compression, i.e., the repulsive part of cell-cell interaction is independent on the type of interacting individuals; while cell-cell adhesion instead depends on the type of the cells involved.
In this respect, dealing with the set of possible explicit form for the interaction kernels proposed in Eq.~\eqref{eq:K-s}, it is consistent to assume that: all interaction kernels have the same behavior at the origin (i.e., for any pair of $p,q\in\{$L,D$\}$ the interaction kernel is characterized by the same values of $s$ and $\FR$), whereas the adhesion strength depends on the type of the interacting cells and it is thereby denoted by $\FA^{pq}$ with $p,q\in\{$L,D$\}$.
The explicit form of the interaction kernels then writes as it follows:
\begin{equation*}
    K^{pq}(r) =
    \left\{
    \begin{array}{ll}
        \ds -\FR\,\left(\dfrac{\dR}{2}\right)^{3-2s}\,r^{2s-3},
            & \hbox{ if } 0 < r \leq \dfrac{\dR}{2}; \\[3mm]
        \ds \dfrac{2\,\FR\,(r-\dR)}{\dR},
            & \hbox{ if } \dfrac{\dR}{2} < r \leq \dR; \\[3mm]
        \ds -\dfrac{4\,\FA^{pq}\,(r-\dA)\,(r-\dR)}{(\dA-\dR)^2},
            & \hbox{ if } \dR < r \leq \dA; \\[3mm]
        0,  & \hbox{ if } r > \dA ,
    \end{array}
    \right.
    \hspace{0.5cm}\textup{with }p,q\in\{\textup{L},\textup{D}\}.
\end{equation*}

\noindent
In order to reproduce the cell sorting, starting from the initial distribution reported in Fig.~\ref{fig:9}, we perform a series of numerical simulations by setting distinct values of both homotypic and heterotypic cell-cell adhesiveness, i.e., $\FA^{pq}$ with $p=q$ and $p\neq q$, respectively; and by taking into account several assumptions about the behavior of the interaction kernels $K^{pq}(r)$, for any $p,q\in\{$L,D$\}$, at the origin.
In this respect, according to previous results, we further state that for any $p,q\in\{$L,D$\}$, the  interaction strengths $\FR$ and $\FA^{pq}$ have to satisfy the H-stability constrain $\FR/\FA^{pq}> F^*$, where $F^*$ depends on the value of $s$ according to Eq.~\eqref{eq:F*s}. In fact, as shown in previous sections, the H-stability of the interaction kernel results fundamental to avoid unrealistic cell overlapping when each cell is represented as a material point.
Entering in more details, according to the numerical simulations performed in previous sections, we here focus only on three possible behaviors of the interaction kernels at the origin: i.e., referring to Eq.~\ref{eq:K-s}, we respectively set $s$ equal to $1.75,\ 1,\ 0.25$, (see Fig.~\ref{fig:10}).
With respect to the interaction strength, according to the results reported in Section~\ref{sec:3}, the value of the repulsive strength $\FR$ is here fixed qual to $100 \unit{\mu m/(\mu g\,s)}$ in all realizations; while the adhesion strengths are set in different ways according to the type of the interacting cells in order to reproduce characteristic dynamics of cell sorting. In all realizations, the value of $\FA^{pq}$, with $p,q\in\{$L,D$\}$, is such that
\begin{equation*}
    \FA^{pq}< \FR/F^* = \left\{
                             \begin{array}{ll}
                               2.68, & \hbox{ if }s=1.75; \\
                               3.16, & \hbox{ if }s=1.0;\\
                               6.51, & \hbox{ if }s=0.25.
                             \end{array}
                           \right.
\end{equation*}
The numerical results reported in the left column in Fig.~\ref{fig:10}, shown that if the heterotypic adhesiveness between the two cell types is higher than the two homotypic adhesiveness interactions (i.e., $\FA^{LD}=\FA^{DL}>\FA^{LL}=\FA^{DD}$), cells heterogeneously mix to form an experimentally observed \emph{checkerboard}.
Conversely, if the homotypic adhesions are stronger than the heterotypic ones (i.e., $\FA^{LL}=\FA^{DD}>\FA^{LD}=\FA^{DL}$, see the central column in Fig.~\ref{fig:10}), we find a spontaneous cell sorting, with the formation of small \emph{clusters} of cells of the same type within the domain.
If further, the adhesion between the light cells is larger than than the heterotypic contact interactions, which is in turn larger than the adhesion between the dark cells (i.e., $\FA^{LL}>\FA^{LD}=\FA^{DL}>\FA^{DD}$, see the right column in Fig.~\ref{fig:10}), we observe the autonomous emergence of little island of light individuals surrounded by a crew of dark cells: this phenomenon is called \emph{engulfment} and was investigated also with other types of models \cite{CPM_GfGa_PRL1992}. Related numerical results were slightly explored in the context of aggregation models in \cite{MKB} without relating them to cell sorting.
From these numerical results, it further emerges that variations in the explicit form of the repulsive part of the interaction kernels, i.e., variations in the value of the parameter $s$, do not significantly affect the evolution of the system.
We find it consistent with the fact that (i) the interaction parameters are always set to guarantee the H-stability of the system, and that (ii) the repulsive strength $\FR$ has been set strong enough to maintain the minimum intercellular distance $d_{\textup{min}}$ greater than the mean cell nucleus $\dN=\dR/2$ (according to previous simulations).

\section*{Conclusions}
\label{concl}

The analysis of cell patterning, as well as the description of the characteristic large-time configurations of cell aggregates, is a relevant issue in developmental biology. The spatial distribution of cells mediates in fact a wide range of physio-pathological phenomena, i.e., from morphogenesis to cancer invasion \cite{Friedl2003,Ilina3203}. In this respect, we have here introduced a discrete microscopic model to reproduce the dynamics of cell systems, where each individual is described by a material point, with concentrated mass, and set to move according to a first-order ODE. In particular, cell velocity has been here set to account for nonlocal adhesive and repulsive contributions, the former including long-range cadherin-mediated mechanisms, the latter modeling cell nucleus resistance to compression.

Both migratory components have been described by introducing a proper family of pairwise interaction kernels, and relative potentials. Such functions are characterized by a negative repulsive part, which may have different slopes, and by a positive parabolic trend in the attractive part. Further, the proposed kernels are intrinsically multiparametric, being determined by a set of free coefficients (also relative to the extension of the interaction regions). The specificity of the cell interaction velocity components obviously impacts on the resulting system behavior, which can range from a dramatic particle collapse to an implausible expansion of the aggregate. However, to properly and realistically reproduce experimental cell patterns, it is necessary to ensure that the model cells reach and maintain typical finite mutual distances.

In this respect, we have here analyzed how the concept of H-stability, derived from statistical mechanics, can relate to the asymptotic behavior of cell particle systems. In particular, our main message is that a crystalline configuration of a cell aggregate can be obtained if the underlying intercellular interaction kernel, and relative potential, satisfies a proper H-stability condition. With this concept in mind, we have then turned to analyze the regions of the space of selected free model parameters (i.e., those concerning cell adhesive and repulsive interactions) that result in the H-stability of a given family of interaction kernels, differing in repulsive part (mainly near the origin). The proposed analytical study has been then enriched by means of numerical realizations that have been able to characterize (in terms of individual spatial configuration and interparticle distance) the large time pattern of cell systems, upon variations also of cell mass and cell number. In this respect, our analysis has also shown that if we aim to derive a continuous macroscopic model from the proposed discrete approach, via coarse-grain procedures, not H-stable potentials must be instead chosen: they in fact allows from minimal intercellular distance converging to zero if $N \rightarrow \infty$ thereby controlling the dimension of the overall cell aggregate.

Our study has also relevant implications in a number of biological applications. First of all, it \textit{a priori} restricts the possible variations of the free interaction parameters, allowing accurate calibrations and estimates without the need of massive preliminary simulations. It is also possible to solve the inverse problem of finding suitable interaction potentials to reproduce a given configuration of an experimental colony. In particular, this application involves optimization issues. Finally, our analysis allows to easily reproduce cell sorting phenomena by only tuning the different adhesiveness of the component populations, provided that the relative sets of parameters result in H-stable systems.

It is however important to notice that the proposed work, as well as the present applications, is based on the assumption that cell behavior is completely determined by adhesive/repulsive contributions. This is of course an oversimplification of the biological picture. Cell migration is in fact a quite complex process involving several other mechanisms and stimuli, such as chemotaxis (i.e., cell locomotion up to gradients of a diffusible chemical field) or durotaxis (i.e., cell locomotion towards stiffer regions of the matrix environment). In this respect, it would be interesting to include in our cell model some of these velocity components and to study how they possibly affect the stable configuration of the particle system. The development of the proposed approach in this respect obviously extends the range of possible biological applications.


\section*{Acknowledgments}
JAC acknowledges support by the Engineering and Physical Sciences Research Council (EPSRC) under grant no. EP/P031587/1, by the Royal Society and the Wolfson Foundation through a Royal Society Wolfson Research Merit Award and by the National Science Foundation (NSF) under grant no. RNMS11-07444 (KI-Net).
AC acknowledge partial funding by the Politecnico di Torino and the Fondazione Cassa di Risparmio di Torino in the context of the funding campaign ``La Ricerca dei Talenti'' (HR Excellence in Research). AC and MS acknowledge Istituto Nazionale di Alta Matematica (INdAM) ``Francesco Severi'' and the ``Gruppo Nazionale per la Fisica Matematica'' (GNFM).


\bibliographystyle{abbrv}
\bibliography{references}

\begin{thebibliography}{10}

\bibitem{Alber2003}
M.~S. Alber, M.~A. Kiskowski, J.~A. Glazier, and Y.~Jiang.
\newblock {\em On Cellular Automaton Approaches to Modeling Biological Cells},
  pages 1--39.
\newblock Springer New York, New York, NY, 2003.

\bibitem{Anderson2007}
A.~Anderson and K.~Rejniak.
\newblock {\em Single-Cell-Based Models in Biology and Medicine}.
\newblock Mathematics and Biosciences in Interaction. Birkh{\"a}user Basel,
  2007.

\bibitem{BCLR2}
D.~Balagu\'e, J.~A. Carrillo, T.~Laurent, and G.~Raoul.
\newblock Dimensionality of local minimizers of the interaction energy.
\newblock {\em Arch. Ration. Mech. Anal.}, 209(3):1055--1088, 2013.

\bibitem{BKSUV}
A.~L. Bertozzi, T.~Kolokolnikov, H.~Sun, D.~Uminsky, and J.~von Brecht.
\newblock Ring patterns and their bifurcations in a nonlocal model of
  biological swarms.
\newblock {\em Commun. Math. Sci.}, 13(4):955--985, 2015.

\bibitem{BV}
M.~Bodnar and J.~J.~L. Vel\'azquez.
\newblock Friction dominated dynamics of interacting particles locally close to
  a crystallographic lattice.
\newblock {\em Math. Methods Appl. Sci.}, 36(10):1206--1228, 2013.

\bibitem{BDFS}
M.~Burger, M.~Di~Francesco, S.~Fagioli, and A.~Stevens.
\newblock Sorting phenomena in a mathematical model for two mutually
  attracting/repelling species.
\newblock {\em preprint arXiv:1704.04179}.

\bibitem{CCP15}
J.~A. Ca\~nizo, J.~A. Carrillo, and F.~S. Patacchini.
\newblock Existence of compactly supported global minimisers for the
  interaction energy.
\newblock {\em Arch. Ration. Mech. Anal.}, 217(3):1197--1217, 2015.

\bibitem{CP16}
J.~A. Ca\~nizo and F.~S. Patacchini.
\newblock Discrete minimisers are close to continuum minimisers for the
  interaction energy.
\newblock {\em preprint arXiv:1612.09233}.

\bibitem{CC}
V.~Calvez and J.~A. Carrillo.
\newblock Volume effects in the {K}eller-{S}egel model: energy estimates
  preventing blow-up.
\newblock {\em J. Math. Pures Appl. (9)}, 86(2):155--175, 2006.

\bibitem{meanfield}
J.~A. Carrillo, Y.-P. Choi, and M.~Hauray.
\newblock The derivation of swarming models: mean-field limit and {W}asserstein
  distances.
\newblock In {\em Collective dynamics from bacteria to crowds}, volume 553 of
  {\em CISM Courses and Lect.}, pages 1--46. Springer, Vienna, 2014.

\bibitem{CDM16}
J.~A. Carrillo, M.~G. Delgadino, and A.~Mellet.
\newblock Regularity of local minimizers of the interaction energy via obstacle
  problems.
\newblock {\em Comm. Math. Phys.}, 343(3):747--781, 2016.

\bibitem{CDP}
J.~A. Carrillo, M.~R. D'Orsogna, and V.~Panferov.
\newblock Double milling in self-propelled swarms from kinetic theory.
\newblock {\em Kinet. Relat. Models}, 2(2):363--378, 2009.

\bibitem{CFTV10}
J.~A. Carrillo, M.~Fornasier, G.~Toscani, and F.~Vecil.
\newblock Particle, kinetic, and hydrodynamic models of swarming.
\newblock In {\em Mathematical modeling of collective behavior in
  socio-economic and life sciences}, pages 297--336. Springer, 2010.

\bibitem{CHM14}
J.~A. Carrillo, Y.~Huang, and S.~Martin.
\newblock Explicit flock solutions for {Q}uasi-{M}orse potentials.
\newblock {\em European J. Appl. Math.}, 25(5):553--578, 2014.

\bibitem{CHS17}
J.~A. Carrillo, Y.~Huang, and M.~Schmidtchen.
\newblock Zoology of a non-local cross-diffusion model for two species.
\newblock {\em preprint arXiv:1705.03320}.

\bibitem{CMP13}
J.~A. Carrillo, S.~Martin, and V.~Panferov.
\newblock A new interaction potential for swarming models.
\newblock {\em Phys. D}, 260:112--126, 2013.

\bibitem{CHDOB07}
Y.-L. Chuang, Y.~R. Huang, M.~R. D'Orsogna, and A.~L. Bertozzi.
\newblock Multi-vehicle flocking: scalability of cooperative control algorithms
  using pairwise potentials.
\newblock In {\em Robotics and Automation, 2007 IEEE International Conference
  on}, pages 2292--2299. IEEE, 2007.

\bibitem{CaSmPl_MMNP2015}
A.~Colombi, M.~Scianna, and L.~Preziosi.
\newblock A measure-theoretic model for collective cell migration and
  aggregation.
\newblock {\em Math. Model. Nat. Phenom.}, 10(1):4--35, 2015.

\bibitem{CaSmPl_JMB2017}
A.~Colombi, M.~Scianna, and L.~Preziosi.
\newblock Coherent modelling switch between pointwise and distributed
  representations of cell aggregates.
\newblock {\em J. Math. Biol.}, 74(4):783--808, 2017.

\bibitem{CaSmTa_JMB2014}
A.~Colombi, M.~Scianna, and A.~Tosin.
\newblock Differentiated cell behavior: a multiscale approach using measure
  theory.
\newblock {\em J. Math. Biol.}, 71(5):1049--1079, 2015.

\bibitem{Deutsch2007}
A.~Deutsch, P.~Maini, and S.~Dormann.
\newblock {\em Cellular Automaton Modeling of Biological Pattern Formation:
  Characterization, Applications, and Analysis}.
\newblock Modeling and Simulation in Science, Engineering and Technology.
  Birkh{\"a}user Boston, 2007.

\bibitem{DTGC}
P.~Domschke, D.~Trucu, A.~Gerisch, and M.~A.~J. Chaplain.
\newblock Mathematical modelling of cancer invasion: implications of cell
  adhesion variability for tumour infiltrative growth patterns.
\newblock {\em J. Theoret. Biol.}, 361:41--60, 2014.

\bibitem{OCBC06}
M.~R. D'Orsogna, Y.-L. Chuang, A.~L. Bertozzi, and L.~S. Chayes.
\newblock Self-propelled particles with soft-core interactions: patterns,
  stability, and collapse.
\newblock {\em Physical review letters}, 96(10):104302, 2006.

\bibitem{Drasdo2003}
D.~Drasdo.
\newblock {\em On Selected Individual-based Approaches to the Dynamics in
  Multicellular Systems}, pages 169--203.
\newblock Birkh{\"a}user Basel, Basel, 2003.

\bibitem{Friedl2003}
P.~Friedl and K.~Wolf.
\newblock Tumour-cell invasion and migration: diversity and escape mechanisms.
\newblock {\em Nat Rev Cancer}, 3(5):362--374, May 2003.

\bibitem{GC}
A.~Gerisch and M.~A.~J. Chaplain.
\newblock Mathematical modelling of cancer cell invasion of tissue: local and
  non-local models and the effect of adhesion.
\newblock {\em J. Theoret. Biol.}, 250(4):684--704, 2008.

\bibitem{CPM_GfGa_PRL1992}
F.~Graner and J.~A. Glazier.
\newblock Simulation of biological cell sorting using a two-dimensional
  extended potts model.
\newblock {\em Phys. Rev. Lett.}, 69:2013--2016, Sep 1992.

\bibitem{Hillen01}
T.~Hillen and K.~Painter.
\newblock Global existence for a parabolic chemotaxis model with prevention of
  overcrowding.
\newblock {\em Adv. in Appl. Math.}, 26(4):280--301, 2001.

\bibitem{HP06}
D.~D. Holm and V.~Putkaradze.
\newblock Formation of clumps and patches in self-aggregation of finite-size
  particles.
\newblock {\em Physica D: Nonlinear Phenomena}, 220(2):183--196, 2006.

\bibitem{Ilina3203}
O.~Ilina and P.~Friedl.
\newblock Mechanisms of collective cell migration at a glance.
\newblock {\em Journal of Cell Science}, 122(18):3203--3208, 2009.

\bibitem{review}
T.~Kolokolnikov, J.~A. Carrillo, A.~Bertozzi, R.~Fetecau, and M.~Lewis.
\newblock Emergent behaviour in multi-particle systems with non-local
  interactions [{E}ditorial].
\newblock {\em Phys. D}, 260:1--4, 2013.

\bibitem{MKB}
A.~Mackey, T.~Kolokolnikov, and A.~L. Bertozzi.
\newblock Two-species particle aggregation and stability of co-dimension one
  solutions.
\newblock {\em Discrete Contin. Dyn. Syst. Ser. B}, 19(5):1411--1436, 2014.

\bibitem{MVO05}
D.~Morale, V.~Capasso, and K.~Oelschl{\"a}ger.
\newblock An interacting particle system modelling aggregation behavior: from
  individuals to populations.
\newblock {\em Journal of mathematical biology}, 50(1):49--66, 2005.

\bibitem{MT15}
H.~Murakawa and H.~Togashi.
\newblock Continuous models for cell--cell adhesion.
\newblock {\em Journal of theoretical biology}, 374:1--12, 2015.

\bibitem{Ol}
K.~Oelschl\"ager.
\newblock Large systems of interacting particles and the porous medium
  equation.
\newblock {\em J. Differential Equations}, 88(2):294--346, 1990.

\bibitem{Hillen02}
K.~Painter and T.~Hillen.
\newblock Volume-filling and quorum-sensing in models for chemosensitive
  movement.
\newblock {\em Can. Appl. Math. Q.}, 10(4):501--543, 2002.

\bibitem{PBSG}
K.~J. Painter, J.~M. Bloomfield, J.~A. Sherratt, and A.~Gerisch.
\newblock A nonlocal model for contact attraction and repulsion in
  heterogeneous cell populations.
\newblock {\em Bull. Math. Biol.}, 77(6):1132--1165, 2015.

\bibitem{ruelle}
D.~Ruelle.
\newblock {\em Statistical mechanics: {R}igorous results}.
\newblock W. A. Benjamin, Inc., New York-Amsterdam, 1969.

\bibitem{SmPl_MMS2012}
M.~Scianna and L.~Preziosi.
\newblock Multiscale developments of the cellular {P}otts model.
\newblock {\em Multiscale Model. Simul.}, 10(2):342--382, 2012.

\bibitem{SKT}
N.~Shigesada, K.~Kawasaki, and E.~Teramoto.
\newblock Spatial segregation of interacting species.
\newblock {\em J. Theoret. Biol.}, 79(1):83--99, 1979.

\bibitem{VS}
A.~Volkening and B.~Sandstede.
\newblock Modelling stripe formation in zebrafish: an agent-based approach.
\newblock {\em Journal of The Royal Society Interface}, 12(112), 2015.

\end{thebibliography}

\end{document}